\documentclass[final,5p,times,authoryear]{elsarticle}

\usepackage{amssymb}
\usepackage{xcolor}
\usepackage{braket}
\usepackage{amsmath}
\usepackage{booktabs}
\usepackage{tikz}
\usepackage{pgfplots}
\pgfplotsset{compat=1.18}
\usepackage{caption}
\usepackage{graphicx}

\makeatletter
\def\s@[email protected]{\discretionary{-}{-}{-}}
\makeatother

\journal{International Journal of Heat and Mass Transfer}

\begin{document}

\begin{frontmatter}

\title{Integrating moment tensor potentials with finite-element modeling for heat transfer prediction in FLiBe-based molten salt systems}

\affiliation[a]{organization={Skolkovo Institute of Science and Technology},
            addressline={Skolkovo Innovation Center, Bolshoy Bulvar, 30}, 
            city={Moscow},
            postcode={143026},
            country={Russia}}

\affiliation[b]{organization={Digital Materials LLC},
            addressline={Odintsovo, Kutuzovskaya str. 4A}, 
            city={Moscow Region},
            postcode={143001},
            country={Russia}}
            
\affiliation[c]{organization={Institute of High Temperature Electrochemistry, Ural Branch of the Russian Academy of Sciences},
            addressline={Academicheskaya Street, 20}, 
            city={Yekaterinburg},
            postcode={620066},
            country={Russia}}   

\affiliation[d]{organization={Ural Federal University},
            addressline={Lenin Avenue, 51}, 
            city={Yekaterinburg},
            postcode={620075},
            country={Russia}}

\affiliation[e]{organization={N.A. Dollezhal Research and Development Institute of Power Engineering JSC},
            addressline={Krasnoselskaya str., 2/8}, 
            city={Moscow},
            postcode={107140},
            country={Russia}} 
            
\affiliation[f]{organization={These authors contributed equally}}

\author[a,b,f]{Mikhail Polovinkin}
\author[c,d,f]{Ksenia Abramova}
\author[c]{Oksana Rahmanova}
\author[c]{Farit Valiev}
\author[c]{Andrey Isakov}
\author[e]{Andrey Goryachikh}
\author[c,d]{Alexander Galashev}
\author[c,d]{Yurii Zaikov}
\author[a,b]{Dmitrii Maksimov}
\author[a,b]{Alexander Shapeev}
\author[a,b]{Nikita Rybin}

\begin{abstract}
Molten fluoride salts are promising heat-transfer media for advanced molten salt reactors (MSRs), where reliable thermophysical property determination is critical for component design and safety. We present an integrated multiscale framework that couples machine-learning-driven atomistic simulations with finite-element (FE) modeling to predict the heat-transfer performance of FLiBe-based salts in a linear heat exchanger. At the atomistic scale, Moment Tensor Potentials (MTPs), actively trained on \textit{ab initio} data, are developed for pure FLiBe (66~mol~\% -- 34~mol~\% and 74~mol~\% -- 26~mol~\%  LiF-BeF$_2$ fractions), FLiBe-LaF$_3$, and FLiBe-UF$_4$. These potentials are used in molecular dynamics simulations to obtain temperature- and composition-dependent transport properties (density, viscosity, thermal conductivity, and isobaric heat capacity), which are mapped as inputs to a three-dimensional FE model of the experimental thermal loop. The FE model with reference (literature) transport properties reproduces the experimental heat-transfer behavior of pure FLiBe 66~mol~\% -- 34~mol~\% to within 10~\% in the laminar regime and 18~\% in the transitional and turbulent regimes, validating the end-to-end pipeline for this composition. The same model with MTP-MD-derived transport properties systematically overestimates the heat-transfer coefficient by 25-28~\%, an offset consistent with the MTP-MD biases on thermal conductivity ($+$20-30~\% relative to experiment) and viscosity ($-$20-25~\%). Applied to the ternary systems FLiBe-LaF$_3$ and FLiBe-UF$_4$ over 0-5~mol\%, the MTP-MD-driven FE model predicts a mean reduction in heat-transfer efficiency of 8-11~\% relative to pure FLiBe, with UF$_4$ exhibiting the strongest effect. The qualitative ordering of the three systems is the more robust result; the absolute value of the 8-11~\% figure is contingent on the MTP accuracy. The framework is complementary to high-temperature experiments and provides a physics-based pathway for the rapid screening of MSR coolant formulations.
\end{abstract}

\begin{graphicalabstract}
\includegraphics[width=\textwidth]{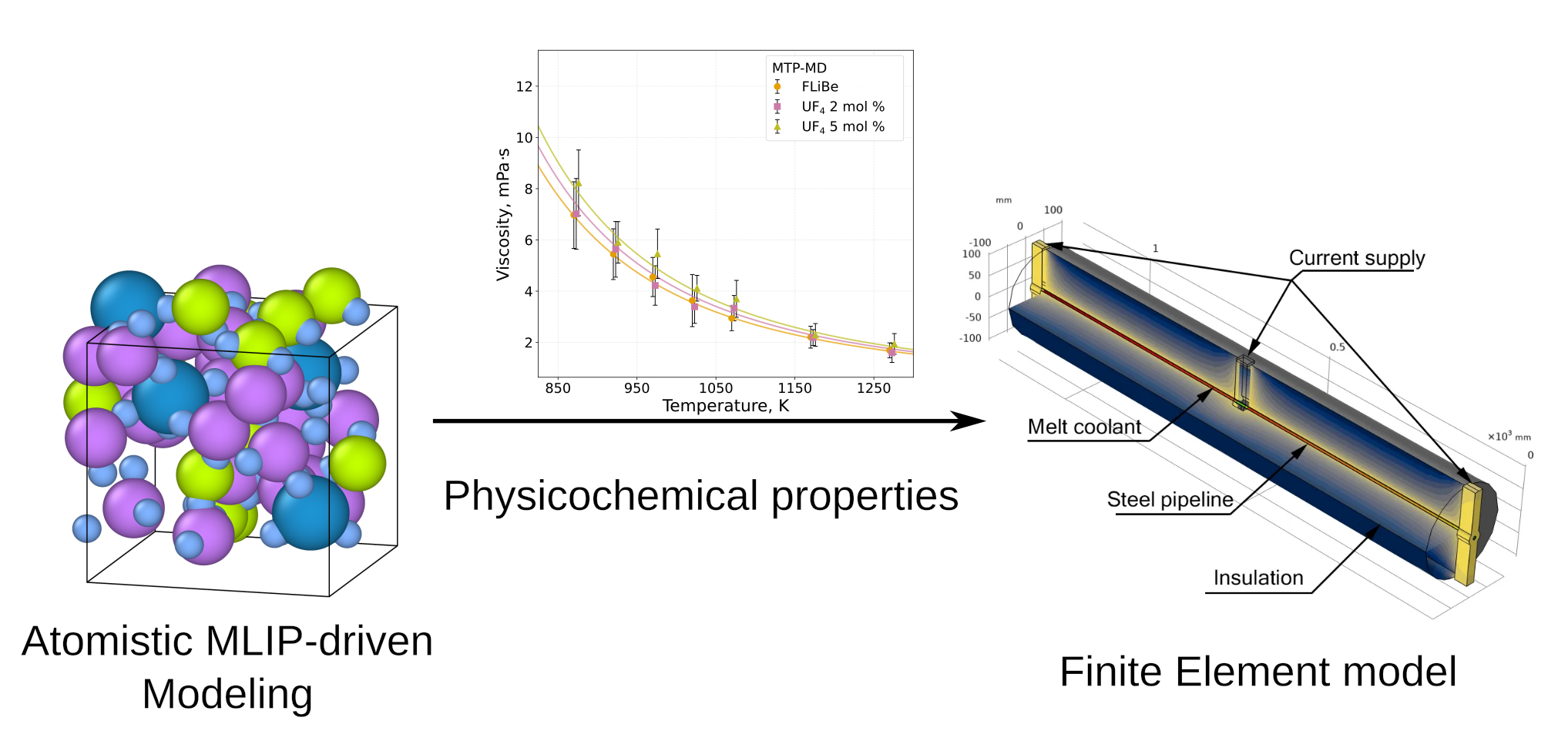}
\end{graphicalabstract}
\onecolumn

\begin{highlights}
\item MTP-MD input bias quantified: viscosity $-$20--25~\% vs experiment, thermal conductivity $+$20--30~\%.
\item Multiscale framework coupling MTP-MD and FE for FLiBe molten-salt heat-transfer.
\item FE pipeline validated for pure FLiBe within 10\% across flow regimes.
\item MTP-MD-driven FE predicts 8--11\% heat-transfer reduction for FLiBe-LaF$_3$/UF$_4$ 4~mol~\%.

\end{highlights}
\twocolumn

\end{frontmatter}

\section{Introduction}
Molten fluoride salts, notably LiF-BeF$_2$ (FLiBe) and LiF-NaF-KF, are pivotal coolants for molten salt reactors (MSRs) owing to their thermal stability and chemical inertness \cite{alberto2022pre, qiu2019numerical, romatoski2017fluoride}. Computational modeling is increasingly indispensable for predicting thermophysical behavior and preventing thermal-hydraulic failures -- including localized crystallization and flow maldistribution -- that compromise MSR operational reliability \cite{leblanc2010molten, ignatiev2014molten, reis2024consequences, deng2026flow}. Accurate knowledge of transport properties (viscosity, thermal conductivity, and heat capacity) underpins such simulations, yet experimental characterization of high-temperature, corrosive melts remains resource-intensive and compositionally constrained \cite{rudenko2022dynamic, serrano2013molten, abou2020radiative}. Recent advances in machine-learned interatomic potentials (MLIPs) have enabled efficient molecular dynamics (MD) simulations of molten salts at near first-principles accuracy, including FLiBe-based systems \cite{errors_Rybin2024, cockrell2025heat, attarian2022thermophysical, li2024compositional, lee2021comparative}.

Despite these atomistic advances, a fully validated pipeline that uses machine-learning interatomic potentials as the sole source of thermophysical inputs and quantifies the resulting systematic offset against an independent experimental thermal loop has not been reported for FLiBe-based systems. The primary goal of this work is to establish a scalable, physics-based computational pathway directly linking atomic-scale interactions to engineering-scale thermal performance, enabling accelerated assessment and optimization of MSR coolant formulations. We present a hierarchically integrated framework that bridges this gap by coupling Moment Tensor Potential (MTP)-driven MD with three-dimensional finite-element (FE) modeling of a linear heat exchanger. MTPs are parameterized for pure FLiBe (66~mol~\% -- 34~mol~\% and 74~mol~\% -- 26~mol~\%, LiF-BeF$_2$ ratios), FLiBe-LaF$_3$, and FLiBe-UF$_4$ melts via fitting to density-functional theory (DFT) data. The resulting temperature- and composition-dependent transport properties are mapped as inputs to a macroscopic FE representation of an experimental thermal loop, advancing prior heat-transfer studies for MSR coolants \cite{srivastava2013heat, balderrama2025cfd, he2022heat}.

At the continuum scale, heat transfer intensity is conventionally quantified by empirical closures -- the Sieder-Tate, Hausen, and Petukhov-Kirillov correlations for laminar, transitional, and turbulent regimes, respectively \cite{sieder1936heat, hausen1959neue, Petukhov1958totheques}. These correlations are calibrated for simple fluids under idealized conditions and do not account for complex salt compositions or variable channel geometries relevant to MSRs. Our FE model overcomes these limitations by resolving local thermophysical fields without empirical closures. The remainder of the paper is organized as follows. Section~\ref{AMR} presents the MTP-MD results for the transport properties of pure FLiBe (66~mol~\% -- 34~mol \% and 74~mol~\% -- 26~mol~\%) and the ternary systems FLiBe--LaF$_3$ and FLiBe--UF$_4$. Section~3.2 validates the FE pipeline against the experimental thermal loop and quantifies the systematic offset introduced when MTP-MD transport properties are used in place of literature values. Section~\ref{sec:fe-md} applies the framework to the ternary systems and predicts the resulting reduction in heat-transfer efficiency. Section~4 summarizes the conclusions and outlines the limitations and future work.

\section{Methods}

\subsection{Atomistic Modeling}

To model the heat transport properties of molten salts, it is necessary to obtain temperature dependencies of density, viscosity, thermal conductivity, heat capacity, diffusion coefficients. All of the aforementioned properties are accessible via the molecular dynamics (MD) simulations. The accuracy of the results is determined by the accuracy of the model used to describe interatomic interactions. Machine-learned interatomic potentials (MLIPs) are currently the most suitable approach for modeling molten salts owing to their accuracy and computational efficiency \cite{moltensalt_Porter2022,shen2025supersalt}. 

\subsubsection{MLIP training}

In this work, we used the Moment Tensor Potential (MTP) to describe interatomic interactions. MTP has been previously used to model various molten salt systems \cite{attarian2022thermophysical,errors_Rybin2024,polovinkin2026machine}, including FLiBe and FLiNaK. Details of the MTP construction and parametrization are briefly described below, and details can be found in original articles \cite{MTP_Shapeev2016,mlip2_Novikov2021,mlip3_Podryabinkin2023}. 

MTP was parametrized using the data (energies, forces acting on atoms, and stresses) obtained from density-functional theory (DFT) calculations. Usually, one performs \textit{ab initio} MD and samples the trajectory to obtain samples for initial training set. In this work, we performed MD simulations with simulations cells containing 100-120 atoms for 300~ps (for each system of interest: FLiBe 66~mol~\% -- 34~mol~\%, FLiBe 66~mol~\% -- 34~mol~\% with LaF$_{3}$ systems with $\sim$2 and $\sim$5~mol~\%, and FLiBe 66~mol~\% -- 34~mol~\% with UF$_{4}$ $\sim$2 and $\sim$5~mol~\% were considered) using a universal MLIP (MACE-MP-0 model \cite{batatia2022mace, batatia2025design}). These trajectories were then sampled uniformly (with 1~ps distance) and collected into two datasets containing 900 samples each: one contains pure FLiBe and FLiBe with LaF$_{3}$, whereas the other one contains pure FLiBe and FLiBe with UF$_{4}$. Universal models like MACE-MP-0 and MatterSim perform only modestly on molten salts due to solid-state training biases. Fine-tuning improves accuracy but rivals the cost of training a dedicated MTP. Given MTP’s comparable accuracy, lower MD cost, and active-learning advantages for stability and robustness, we selected it for production. Universal models are retained solely for initial training-set generation, leveraging their transferability and efficiency.

Energies, forces acting on atoms, and stresses for each structure from the aforementioned sets were then computed using DFT. These data constitute the initial training sets. DFT calculations were performed with the Vienna \textit{ab initio} Simulations Package (VASP version 5.4.4) \cite{Vasp_Kresse1999} with the Perdew-Burke-Ernzerhof (PBE) exchange-correlation functional and projector-augmented wave (PAW) method \cite{PBE_GGA_Perdew1996}. A plane-wave cutoff energy of 600~eV was used in all calculations. Brillouin-zone sampling was restricted to the Gamma-point only (1$\times$1$\times$1 k‑point mesh). Electronic occupancies were smeared using the Gaussian method with a smearing width of 0.05~eV. For systems containing elements with localized f-electrons, the DFT+U method is applied with a Hubbard U implemented in VASP~\cite{bengone2000implementation, rohrbach2003electronic}.

Given that FLiBe is an ionic compound, accounting for long-range interactions is crucial. Previously we have shown \cite{errors_Rybin2024} MTP trained on data including the DFT-D3 correction \cite{grimme2010consistent, grimme2011effect} yields densities in excellent agreement with experiment, while MTP trained without the D3 correction yields a 10~\% error. The functional form of MTP implemented in the MLIP-2 package currently lacks explicit terms for dispersion corrections. Based on our prior experience \cite{errors_Rybin2024,polovinkin2026machine}, no substantial improvements in energy and force convergence are expected beyond a cutoff radius of 5~$\AA$. This distance is sufficient to include several neighboring coordination shells in the liquid, as electrostatic interactions beyond this distance are negligible for FLiBe configurations due to effective screening. Although incorporating long-range interactions directly into the potential construction could, in principle, enhance simulation accuracy, such functionality has not yet been implemented in the current MTP framework. Therefore, we explicitly account for long-range dispersion interactions by employing the D3 correction in DFT calculations.

We fitted (based on the Broyden-Fletcher-Goldfarb-Shanno method) MTPs of level 16, containing 608 fitting parameters, on initial training sets. The influence of different MTP levels (i.e., different numbers of parameters) on the prediction of targeted properties molten salt have been previously investigated for FLiNaK and FLiBe \cite{errors_Rybin2024, attarian2022thermophysical}. Those studies showed that beyond level 16, increasing the MTP complexity does not substantially improve agreement with experimental results, whereas computational costs increases significantly. Hence, for all subsequent calculations in this study, we employed the potential with level 16. 

Low errors do not guarantee the robustness of the potential, i.e., its ability to be used in large-scale and long-timescale MD simulations, since the potential is fitted on a limited set of structures representing the potential energy surface (PES) of the system and may substantially extrapolate when applied to samples from regions of the PES that were not represented in the training set. To develop an accurate and robust potential one can use active learning (AL) procedure, which not only ensures robustness, but also allows one to use minimal amount of data. In this work, we used an AL procedure based on the D-optimality criterion coupled to MaxVol algorithm~\cite{AL_Podryabinkin2017}. During the AL step we run a set of MD trajectories and at each step of the MD we evaluate the extrapolative behavior of the MTP. Once structures on which the MTP heavily extrapolates are found, the MD simulation is halted and these structures are added to the training set. This is done in an iterative manner till no structures with high extrapolation grades are found in 150~ps MD trajectories. We also note that during AL procedure we performed MTP-MD simulations with structures containing about 8 and 12~mol~\% of LaF$_3$ or UF$_4$. The resulting fitting errors are reported in Tab.~\ref{tab:errors}, MTP vs DFT correlation plots for energies per structure and forces acting on atoms are reported in the SI. The resulting MTPs are accurate and robust; i.e., these can be used to perform large scale MTP-MD simulations in order to obtain physicochemical properties. 

\begin{table}[htbp]
\centering
\caption{MTP fitting errors for FLiBe-UF$_4$ and FLiBe-LaF$_3$ systems. The energy error is per training configuration (100--120 atoms); the force error is per atom per Cartesian component. RMSEs are reported in both cases.}
\begin{tabular}{|c|c|c|}
\hline
Addition & UF$_4$ & LaF$_3$ \\ \hline
Energy, eV      & 0.4    & 0.2     \\ \hline
Force, eV/$\AA$ & 93     & 74 \\
\hline
\end{tabular}
\label{tab:errors}
\end{table}

\subsubsection{Physicochemical properties calculations}

We computed the temperature dependencies of the targeted properties of the molten salts FLiBe-UF$_4$ and FLiBe-LaF$_3$. At each state point (temperature-composition pair), we used 10 independent MD simulations in the NPT ensemble at ambient pressure (1 bar) using the Nose-Hoover thermostat and barostat~\cite{Thermo_Nose2002} with a 1~fs time step. Each trajectory was 1~ns long. Density was obtained by sampling the trajectories every 25~ps at each temperature. Viscosity was obtained using the Green-Kubo approach~\cite{GK_Kubo1957, GK_Green1954, GK_Hess2002, GK_Allen2017, GK_Maginn2018} with an integration time of 25~ps, consistent with our convergence tests (reported in the SI) and with previous studies on molten salts \cite{attarian2022thermophysical, errors_Rybin2024}. The diffusion coefficients were calculated with the Einstein method \cite{GK_Maginn2018, moltensalt_Porter2022}. Standard errors were estimated from the independent MD trajectories and are reported as the uncertainty of the computed properties.

Thermal conductivity was computed via the non-equilibrium Muller-Plathe method~\cite{MP_method1997} by imposing a heat flux and measuring the resulting temperature gradient in the NVE ensemble. It was demonstrated in Ref.~\cite{pan2021finite} that this approach yields heat transport properties in molten salts in better agreement with experimental values than the Green-Kubo method. Contrary to the conventional non-equilibrium molecular dynamics (NEMD) approach, which imposes a temperature gradient on the system and observes the resultant heat flux, the reverse NEMD method imposes a heat flux and measures the resultant temperature gradient. In the reverse NEMD algorithm, a temperature gradient is generated by periodically exchanging kinetic energy between two atoms located in different regions. Specifically, if a periodic simulation box is divided into $\textit{N}$ bins (where \textit{N} is an even number) along a given direction (for example \textit{z}), the atoms with the highest kinetic energy in the first bin and the atoms with the lowest kinetic energy in the \textit{N}/2+1-th bin are identified to exchange kinetic energy at regular intervals (ensuring kinetic energy conservation, even with differing atomic masses). This procedure eventually establishes two symmetric temperature gradients. Given that heat flux can be precisely calculated based on the exchanged kinetic energies, thermal conductivity $\kappa$ can then be determined, using the following equation:
\[
    \kappa = -\frac{Q}{2tl_{x}l_{y}dT/dz},
\]
where $Q$ is the total kinetic energy exchanged during a time period $t$, and $l_{x}l_{y}$ is the cross-sectional area perpendicular to $z$ direction. The factor 2 in denominator accounts for the periodicity of the simulation box. 

For thermal conductivity calculations we used a time step of 1~fs and the supercell with 3136--3392 atoms (2.2~nm~$\times$~2.2~nm~$\times$ 8.8~nm) and a kinetic energy swap rate of 20 every 100~steps. For each temperature, after the initial equilibration in the NPT ensemble for 50~ps, production runs of 250~ps were performed in the NVE ensemble. At every simulated state point the results were averaged over 10 independent MD trajectories.

\subsection{Macroscopic modeling using finite-elements method}
At the continuum scale, convective heat transfer intensity can be quantified by the heat transfer coefficient, governed by the similarity relation
\begin{equation}
    Nu = f(Re, Pr)
    \label{eq:Nu(Re)}
\end{equation}
where $Nu$, $Re$, and $Pr$ are the Nusselt, Reynolds, and Prandtl numbers. Standard engineering practice relies on empirical closures --- the Sieder-Tate correlation for laminar flow ($Re<2300$) \cite{sieder1936heat}, Hausen \cite{hausen1959neue}, Petukhov-Kirillov \cite{Petukhov1958totheques} correlations  for transitional and turbulent flow regimes ($Re>2300$). While convenient, these correlations are calibrated for simple fluids under idealized smooth-channel conditions and do not account for complex salt compositions, variable cross-sections, or coupled forced-natural convection --- all relevant to MSR heat exchangers.

To overcome these limitations, our FE model resolves local thermophysical fields without relying on global empirical closures, enabling detailed, composition-dependent performance assessment. 

A mathematical model describing the thermophysical processes occurring in the thermal test facility is developed based on the conditions and parameters of a real experimental thermal facility. Below is a description of the thermal facility's operation and experimental conditions, along with details of the developed mathematical model of the digital twin of thermal facility. The finite-element model is developed using COMSOL Multiphysics 6.0 \cite{comsol60}.

The model thermal test facility consists of a thin cylindrical steel pipe (heat exchanger) through which the heat carrier, FLiBe eutectic molten salt (66~mol~\% -- 34~mol~\%), flows. The heat exchanger is surrounded by an insulation layer. Heating is achieved by Joule heating via current supplies connected to opposite ends of the heat exchanger, with the central current supply serving as a current collector. The heat carrier, at a specified flow rate and a defined temperature, is supplied to the inlet of the working section. The flow rate and the magnitude of the heating power are determined based on the required flow regime: laminar, transitional, or turbulent.

The materials of the components of the thermal test facility, as well as their geometric characteristics, are shown in Table~\ref{tab:materials_for_FEM}. 

\begin{table}[htbp]
\centering
\caption{The materials used in the simulation of a thermal test facility for studying the heat transfer characteristics of molten FLiBe (66~mol~\% -- 34~mol~\%) with a heat exchanger wall (*$d$ is the diameter, $L$ is the length, $H$ is the height)}
\label{tab:materials_for_FEM}
\resizebox{\columnwidth}{!}{%
\begin{tabular}{llc}
\hline
\textbf{Node} & \textbf{Material} & \textbf{Characteristic size* (mm)} \\
\hline
Heat exchanger (pipe) & Steel AISI 316L & 
    \begin{tabular}[t]{@{}l@{}} 
    \(d = 6\) \\ 
    \(L = 1200\)
    \end{tabular} \\

Thermal insulation of the heat exchanger & 
    \begin{tabular}[t]{@{}l@{}} 
    Kaolin wool \\
    50Al\textsubscript{2}O\textsubscript{3} - 50SiO\textsubscript{2} \\
    (0.128 g/cm\textsuperscript{3})
    \end{tabular} & 
    \begin{tabular}[t]{@{}l@{}} 
    \(d = 200\) \\ 
    \(L = 1200\)
    \end{tabular} \\

Current supply & Steel AISI 321 & H = 240 \\
\hline
\end{tabular}%
}
\end{table}

\begin{figure}[h]
    \centering
    \includegraphics[width=1\linewidth]{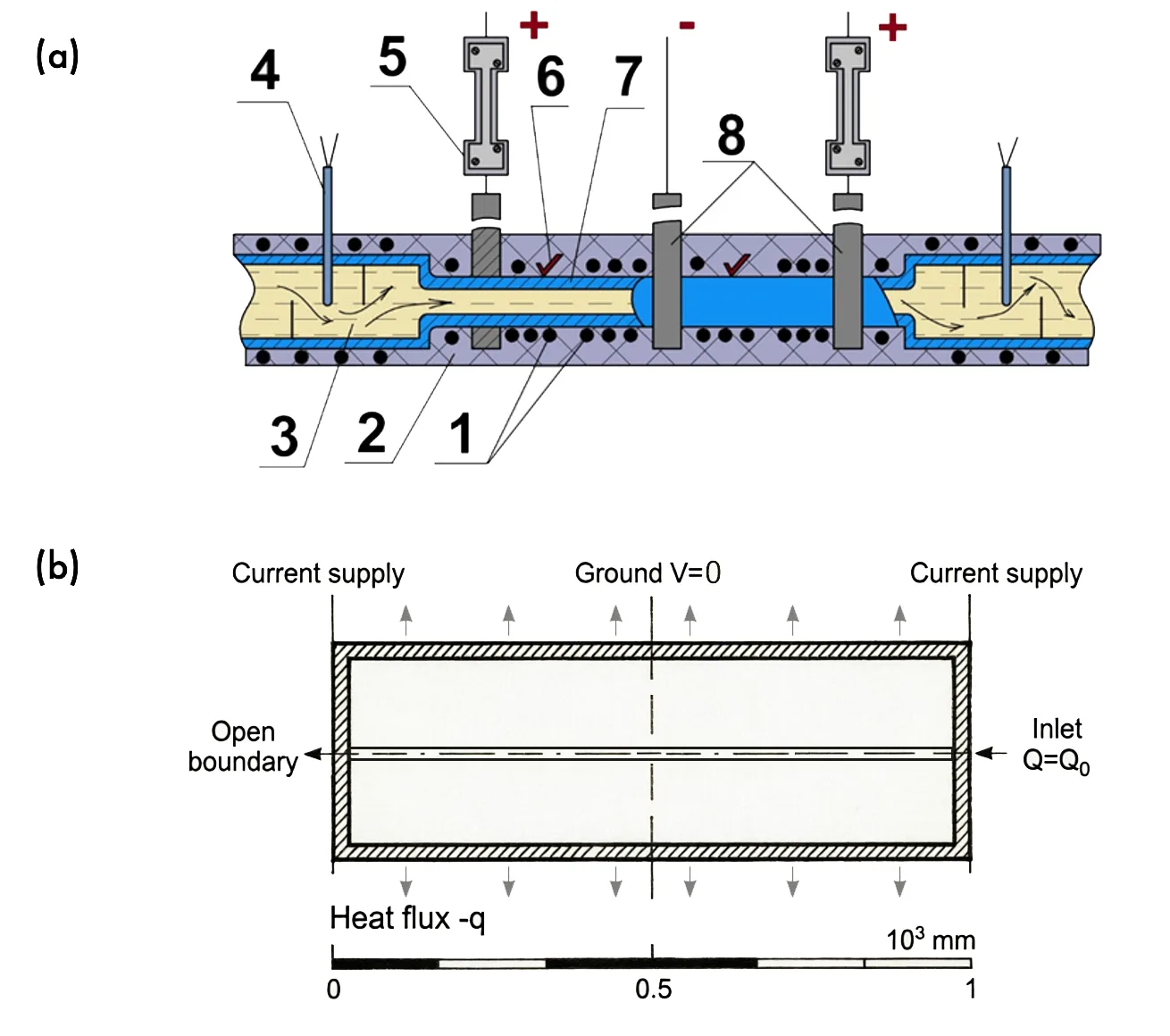}
    \caption{The schematic diagrams of (a) the experimental working section: 1 -- resistance heaters; 2 -- thermal insulation; 3 -- mixing chamber; 4 -- thermocouple; 5 -- current-measuring shunt; 6 -- thermocouple mounting surface; 7 -- working section (nominal diameter 6 mm); 8 -- current leads, and (b) the model with boundary conditions applied.}
    \label{fig:scheme}
\end{figure}

Fig. \ref{fig:scheme} (a) shows the full-scale computational model with predefined boundary conditions. The electrical model was configured using known experimentally voltage drop $V_o$  (Table \ref{tab:regimes}). The electrical current flowed from the outer electrodes toward the central one. Additionally, electrical insulation conditions are specified on all external surfaces of the model.

The voltage drop $V_o$ and the mass flow rate $Q_o$ of the coolant are the initial (input) data for the simulation. The convective heat flux boundary condition is specified on the external surfaces of the insulation: 

\begin{equation}
   -\vec n \cdot \vec q = \alpha \left( T_{amb} - T_i \right),
    \label{eq:4}
\end{equation}

where $\vec n \cdot \vec q$ is the heat flux vector from the wall in the $\vec n$ direction, $\alpha$ is the heat transfer coefficient for natural convection cooling of the insulation wall (W/(m$^2\cdot$K)), and $T_{amb}$ = 300~K is the ambient temperature; $T_i$ is the temperature calculated on the insulation surface (K). 

The melt is supplied through the right boundary (Inlet). The melt flow velocity is determined by the mass flow rate for each regime under investigation (Table \ref{tab:regimes}). The left boundary has a zero excess pressure condition and functions as a "sink." To describe the flow of the melt through the pipe, the Navier-Stokes equations are solved in a laminar regime (500 < Re < 1400) or using a turbulent approximation (2300 < Re < 11000). 

To reflect the actual construction of the facility, the Open Boundary condition is defined for the melt at the outlet of the linear section. It describes the heat flux, and its intensity is calculated as follows:

\begin{equation}
   \Delta H \;=\; \int_{T_{upstr}}^{T} C_p \, dT,
    \label{eq:5}
\end{equation}

where $\Delta H$ is the enthalpy of the flow (J/kg), $C_p$ is the isobaric heat capacity (J/kg$\cdot$K), and $T_{upstr}$ is the temperature of the upstream flow beyond the boundary at which the condition is defined (K). The complete FE-model details, including mathematical model, boundary condition application details and description of the computational mesh development are presented in Supplementary A.

\subsubsection{Validation metrics}
\label{subsubsec:fe_metrics}

The validation of the finite-element model against the experiment is based on three quantities: the outer-wall temperature profile, the local heat-transfer coefficient, and the integrated $Nu$--$Re$ criterion correlation. The local heat-transfer coefficient is defined as
\begin{equation}
  \alpha_{\rm loc} = \frac{q}{T_{\rm wall}(y) - T_{\rm avg}},
    \label{eq:alpha_loc}
\end{equation}
where $q$ is the FE-computed local heat flux normal to the inner steel surface (W/m$^2$), $T_{\rm wall}(y)$ is the local inner-wall temperature (K), and $T_{\rm avg}$ is the local mass-flux-weighted cross-sectional mean bulk temperature (K). 

The Nusselt, Reynolds, and Prandtl numbers used in the criterion correlations are defined as
\begin{subequations}
\label{eq:dimensionless}
\begin{align}
Re  &= \frac{\rho\, U\, D}{\mu},
\label{eq:Re} \\[2pt]
Pr  &= \frac{\mu\, C_p}{k},
\label{eq:Pr} \\[2pt]
Nu  &= \frac{\alpha_{\rm loc}\, D}{k},
\label{eq:Nu}
\end{align}
\end{subequations}
\noindent where $D = 6$~mm is the inner pipe diameter, $U$ is the area-averaged axial velocity, $\alpha_{\rm loc}$ is the local heat-transfer coefficient, and $\rho$, $\mu$, $k$, $C_p$ are the salt density, dynamic viscosity, thermal conductivity, and isobaric heat capacity, all evaluated at the local bulk temperature.

\subsection{Experimental setup}
\label{subsec:exp}

The working section was insulated with 100~mm thick kaolin wool. A schematic diagram of the test section, including sensor locations, is presented in Fig.~\ref{fig:scheme} (a).

The experimental thermal test facility was  instrumented with eighteen thermoelectric transducers (OWEN N 444-09.200/5.0K.1 resistive thermometers), were deployed, with two at the inlet and two at the outlet of the test section for bulk-fluid temperature measurement, and fourteen distributed along the working section for outer-wall temperature measurement (Figure \ref{fig:scheme} a).

Before each measurement, the setup was conditioned to minimize environmental influences on the accuracy of the measurements. The mixing chambers and current leads were independently preheated to temperatures between 823 and 1023~K in 25~K increments. The supplied power was verified by voltage and current measurements using multimeters or a wattmeter. Direct current heating of the channel employs the tube itself as a resistive heating element, which minimizes heat losses.

Heat losses to the environment were determined in a separate calibration run by heating the test section, while purging the tube interior with an inert atmosphere. For each temperature setpoint, the system was stabilized for one hour, after which the steady-state power required to maintain the target temperature was recorded. This recorded power was taken as the measure of heat losses, thereby accounting for parasitic heat exchange from the tube surface. 

During each experimental run, the working temperature was set, and melt flow was established through the model tube (length: 1200~mm , inner diameter: 6~mm) simultaneously with DC power application. The parameters that distinguish each experimental run are given in Table \ref{tab:regimes}.

\begin{table}[htbp]
\centering
\footnotesize
\caption{Initial data for the model thermal test facility (obtained from experiment): voltage drop $V_o$ and mass flow rate $Q$ of the coolant FLiBe}
\begin{tabular}{ccccc}
\toprule
\textbf{Regime} & \textbf{Voltage drop} & \textbf{Mass flow rate} & \textbf{\(T_{input}\) (K)} \\
    & \(V_0,{exp}\) (V) & \(Q_0,{exp}\) (kg/s) &   \\
\midrule
\multicolumn{4}{c}{Laminar flow} \\
1 & 4.31 & 0.016 & 933 \\
2 & 5.33 & 0.031 & 933 \\
\midrule
\multicolumn{4}{c}{Turbulent flow} \\
3 & 7.66 & 0.11  & 893 \\
4 & 9.14 & 0.117 & 893 \\
5 & 9.36 & 0.195 & 893 \\
6 & 12.50 & 0.277 & 893 \\
\bottomrule
\end{tabular}
\label{tab:regimes}
\end{table}

The heat-transfer coefficient was derived from the inlet-to-outlet temperature difference and the net power supplied.Direct measurement of the outer pipeline wall temperature made it possible to determine the local inner wall temperature $T_{wall}(y)$:

\begin{equation}
    T_{wall}(y) = T_{wall}^{outer,exp}(y) - \frac{q \delta_{wall}}{k_{steel}},
    \label{eq:Texp}
\end{equation}

where $q$ is the heat flux density on the inner surface of the pipe (W/m$^2$); $\delta_{\rm wall}$ = 1~mm, is the steel wall thickness.

Current was recorded with a 75SMP1-50-0.5 shunt. Electrical power, inlet temperature, outlet temperature, and wall temperature were registered, with the measured power subsequently corrected for heat losses. Electrical parameters were recorded using a multimeter Appa 506B; voltage was measured at the current-supply contacts, while tube current was determined from the voltage drop across a measuring shunt. Heating current was similarly derived using a stationary shunt.

Volumetric flow rate at the working-section inlet was determined from the time-dependent melt-level variation in the reservoir: the slope of the linear level-versus-time trend was converted to flow rate as the product of the vessel cross-sectional area and the rate of level change. These values were cross-validated against the heat-balance-derived flow rate according to:
    
\begin{equation}
    G_T = \frac{Q_w - Q_s}{C_p \, \bar{\rho} \, (T_{out} - T_{in})},
    \label{eq:flow_rate}
\end{equation}
    
where $G_T$ is the volumetric flow rate (m$^3$/s); $Q_w$ is the electric power supplied to the working section, corrected for heat losses (W); $Q_s$ is the power corresponding to heat losses (W); $C_p$ is the specific heat capacity of the melt (J/(kg$\cdot$K)); $\bar{\rho}$ is the mean density evaluated at the arithmetic average of inlet and outlet temperatures (kg/m$^3$); and $T_{out}$ and $T_{in}$ are the mean flow temperatures at the outlet and inlet (K), respectively.

The heat transfer parameters were determined based on the results of thermal rig tests using the reference temperature-de\-pen\-dent properties of the FLiBe base coolant. These parameters were subsequently used in the development of the finite-element validation model.
The corresponding data are presented in Supplementary Materials.

\section{Results and Discussion}

\subsection{Atomistic Modeling Results}
\label{AMR}

In this section we discuss the MTP-MD results in the following order: local structure (RDF and coordination clusters), density, viscosity, diffusion coefficients, and thermal conductivity.

\begin{figure}[h]
	\centering 
	\includegraphics[trim={0cm 0cm 0cm 0cm}, clip, width=1\linewidth]{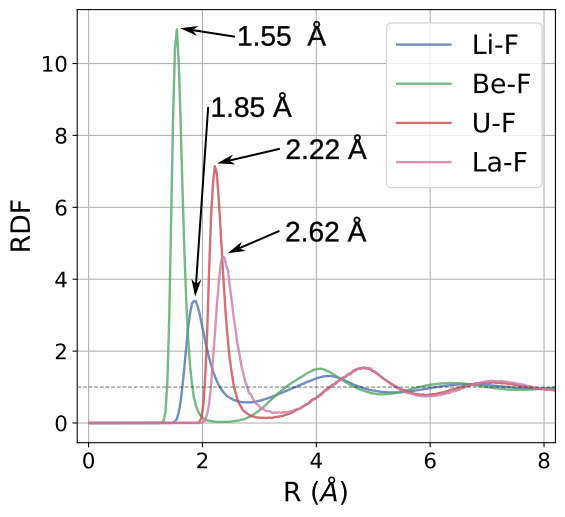}	
	\caption{Radial distribution functions in FLiBe-UF$_4$ (2~mol~\%) and FLiBe-LaF$_3$ (2~mol~\%), extracted from MTP-MD at temperature T=973~K for F-Li, F-Be, F-U, and F-La ion pairs.} 
	\label{fig:rdf}
\end{figure}

\begin{figure}[h]
	\centering 
	\includegraphics[trim={0cm 0cm 0cm 0cm}, clip, width=1\linewidth]{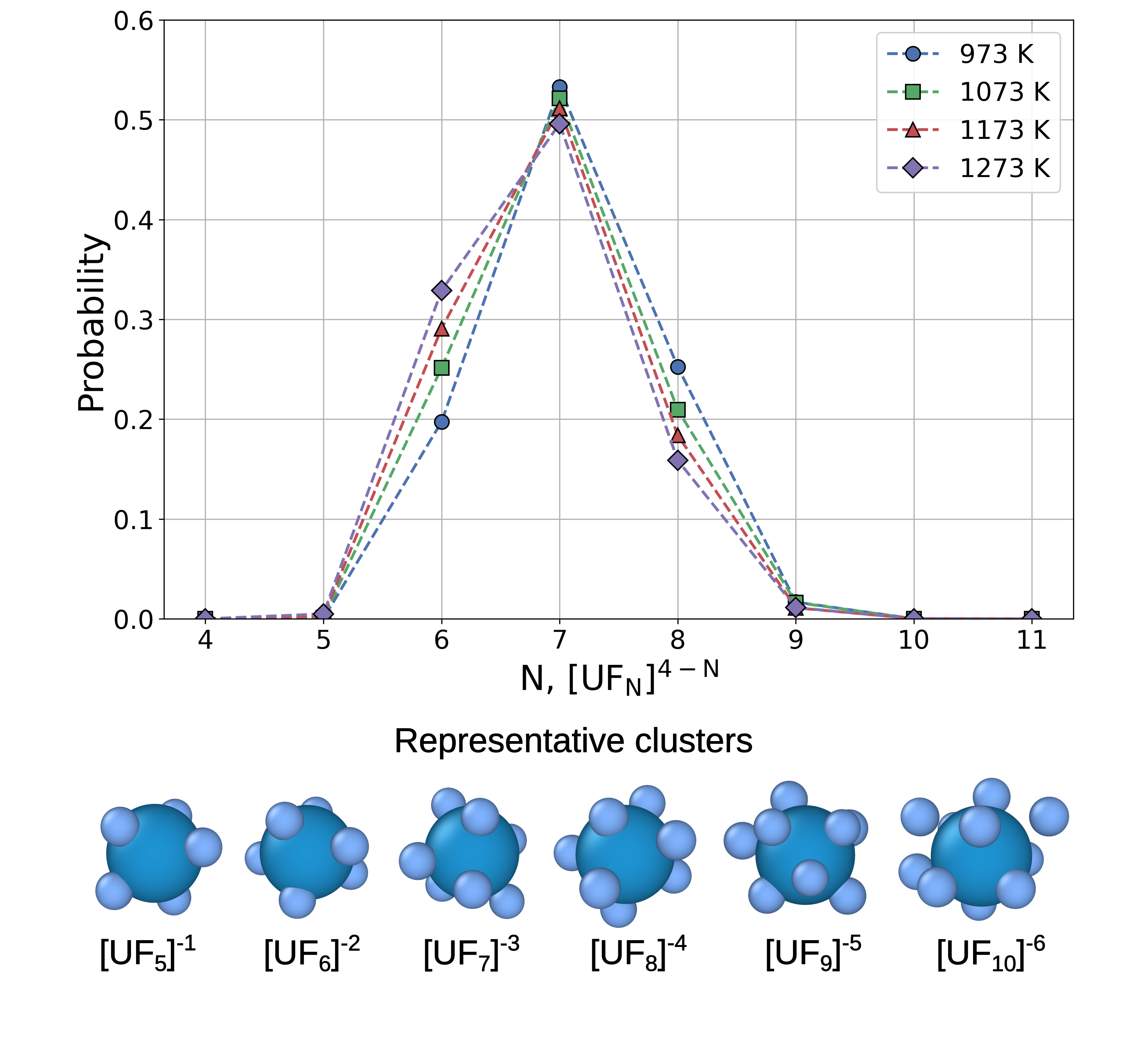}	
	\caption{Distribution of clusters of UF in FLiBe-UF$_4$ (2~mol~\%) computed at various temperatures.} 
	\label{fig:clusters}
\end{figure}

To assess how well the developed MTP can predict local structural features of molten FLiBe-UF$_4$ and FLiBe-LaF$_3$, we first calculated radial distribution functions (RDF). For MTP-MD simulations, the simulation cell contained 824 (FLiBe-UF$_4$ 2~mol~\%) and 816 atoms (FLiBe-LaF$_3$ 2~mol~\%). The use of larger supercells improves the statistical quality of the computed properties and MTP-MD allows performing long simulations, which are inaccessible by AIMD. RDFs for Li-F, Be-F, U-F, and La-F are shown in Fig.~\ref{fig:rdf}. RDFs for Li-F and Be-F pairs are similar for both FLiBe-UF$_4$ and FLiBe-LaF$_3$. Li-F RDF exhibits the typical behavior of liquids with the first peak at 1.85~$\AA$. RDF for Be-F shows a sharp peak at 1.55~$\AA$ and then rapidly decays toward zero. This suggests a strong bonding between Be and F. On the other hand, the RDFs for U-F and La-F have a wider first peak at 2.22~$\AA$ and 2.62~$\AA$, respectively. This result suggests that the first nearest neighbor F shell for U and La is more diffusive and not as well defined as for Be-F, which further indicated that the bonding between U and La and F is not as strong as between Be and F. However, U coordinates F stronger than La does. To further assess the interactions between U and F, we analyzed the statistical distribution of U-F coordination clusters, as shown in Fig.~\ref{fig:clusters}. Mostly the system contains [UF$_{6}$]$^{-2}$, [UF$_{7}$]$^{-3}$, [UF$_{8}$]$^{-4}$, with approximately half of the U-F clusters adopting the [UF$_{7}$]$^{-3}$ form. The portion of [UF$_{6}$]$^{-2}$ changes from roughly 0.33 to 0.2 over the temperature range from 973 to 1273~K. At the same time the portion of [UF$_{8}$]$^{-4}$ clusters grows from 0.17 to 0.25 in the same temperature range. Notably, a small fraction of [UF$_{9}$]$^{-5}$ and [UF$_{10}$]$^{-6}$ clusters is present. The average coordination numbers at temperatures of 973~K and 1273~K are 7.09 and 6.84, respectively. The obtained structural properties are in good agreement with a previous AIMD study \cite{Li_FLiBe_RDF}.

Having established the local structure, we now turn to the temperature dependence of the density. While experimental measurements are feasible for pure FLiBe, computational modeling enables rapid and cost-effective exploration of various salt mixtures with diverse components and compositions, for example, FLiBe-LaF$_3$ and FLiBe-UF$_4$. Performing experiments with the latter is challenging. The density of molten salts has been experimentally measured over a range of temperatures \cite{redkin2021density,rudenko2024density,vidrio2022density,gardner2025effect}, and recent efforts have employed MLIPs based on neural networks and MTP to calculate this property \cite{moltensalt_Porter2022,pan2026machine,rodriguez2021thermodynamic}.

\begin{figure*}
	\centering 
	\includegraphics[trim={0cm 0cm 0cm 0cm}, clip, width=1\linewidth]{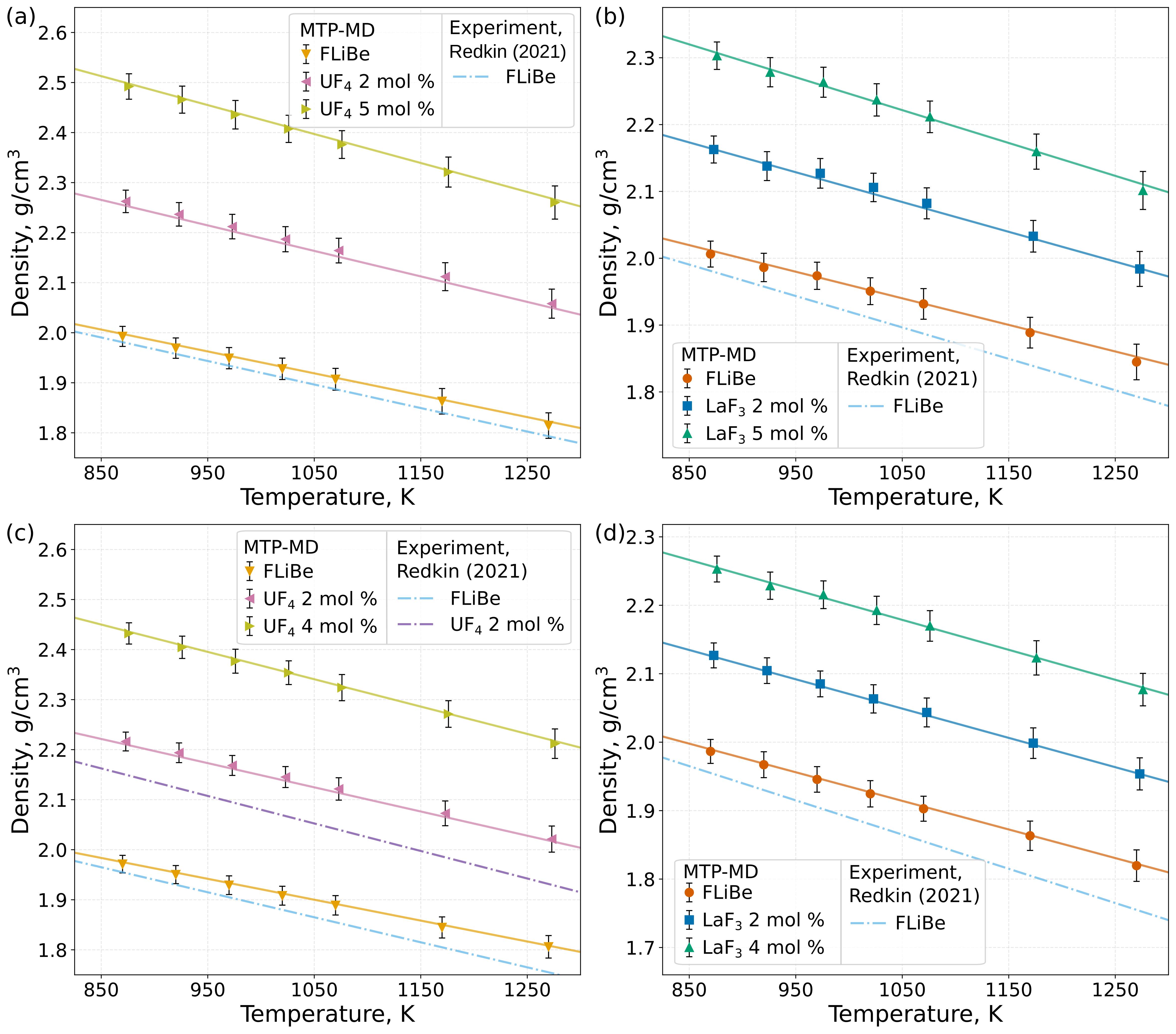}	
	\caption{Temperature dependence of the density of molten salts, obtained with MTP-MD. (a) FLiBe (66~mol~\% -- 34~mol~\%)-UF$_4$, (b) FLiBe (66~mol~\% -- 34~mol~\%)-LaF$_3$, (c) FLiBe (74~mol~\% -- 26~mol~\%)-UF$_4$, and (d) FLiBe (74~mol~\% -- 26~mol~\%)-LaF$_3$. Comparison with the experimantal data from Redkin (2021) \cite{redkin2021density} is presented by dash-dotted lines.} 
	\label{fig:density}
\end{figure*}

We perform density calculations of molten FLiBe (also with LaF$_3$ and UF$_4$ additions) using the same simulation cell as for the structural analysis (comprising 784-864 atoms for FLiBe with 0-5~mol~\% of additions). The results, shown in Fig.~\ref{fig:density}, are in good agreement with several experimental results \cite{redkin2021density}, despite not fixing experimentally-known density during the preparation of the training set. The relative errors for the pure FLiBe salt are 1-4~\%. The density of both FliBe-LaF$_3$ and FLiBe-UF$_4$ molten salts linearly increases with increasing additive concentration. The effect of UF$_4$ addition is more significant compared to that of LaF$_3$. Addition of 2~mol~\% UF$_4$ increases the density of FLiBe 66~mol~\% -- 34~mol~\% by 13~\%, while LaF$_3$ only by 7~\% at temperature 923~K. Both compositions of pure FLiBe 66~mol~\% -- 34~mol~\% and 74~mol~\% -- 26~mol~\% exhibit similar density in our calculations. However, the density increases more substantially for the FLiBe 66~mol~\% -- 34~mol~\% composition upon increasing the additive concentration.

To enable extrapolation beyond the simulated state points for the temperature- and composition-dependent density, $\rho$, we use a linear equation with composition-dependent coefficients:
\begin{equation}
    \rho(T, m) = (\rho_0+\alpha \cdot m) + (b+\beta \cdot m)\cdot T,
    \label{eq:rho_app}
\end{equation}
where $T$ is the temperature, $m$ is the molar fraction of the additive (LaF$_3$ or UF$_4$), $\rho_0$ and $b$ are the zero-temperature intercept (a fitting parameter, not a physical density at 0~K) and the temperature coefficient for pure FLiBe, $\alpha$ and $\beta$ are coefficients capturing the linear variation of the corresponding coefficients with composition. The fitting coefficients are listed in Tab.~\ref{tab:dens}.

\begin{table}[h]
\centering
\caption{Coefficients for density calculation according to equations \ref{eq:rho_app} for systems computed in our research. }
\begin{tabular}{|c|cc|cc|}
\hline
Base salt LiF--BeF$_2$                     & \multicolumn{2}{c|}{66--34}                & \multicolumn{2}{c|}{74--26}               \\ \hline
Addition                      & \multicolumn{1}{c|}{La}           & U            & \multicolumn{1}{c|}{La}           & U            \\ \hline
$\rho_0$, g/cm$^3$            & \multicolumn{1}{c|}{2.38}       & 2.38       & \multicolumn{1}{c|}{2.35}       & 2.33       \\ \hline
$\alpha$, g/cm$^3$            & \multicolumn{1}{c|}{8.75}       & 13.66      & \multicolumn{1}{c|}{8.17}       & 15.24      \\ \hline
$b\cdot 10^{4}$ , g/(cm$^3 \cdot$K) & \multicolumn{1}{c|}{-4.24} & -4.44 & \multicolumn{1}{c|}{-4.16} & -4.15 \\ \hline
$\beta\cdot 10^{3}$, g/(cm$^3 \cdot$K)   & \multicolumn{1}{c|}{-2.40} & -3.00 & \multicolumn{1}{c|}{-1.16} & -3.58 \\ \hline
\end{tabular}
\label{tab:dens}
\end{table}

\begin{figure*}
	\centering 
	\includegraphics[width=1\linewidth]{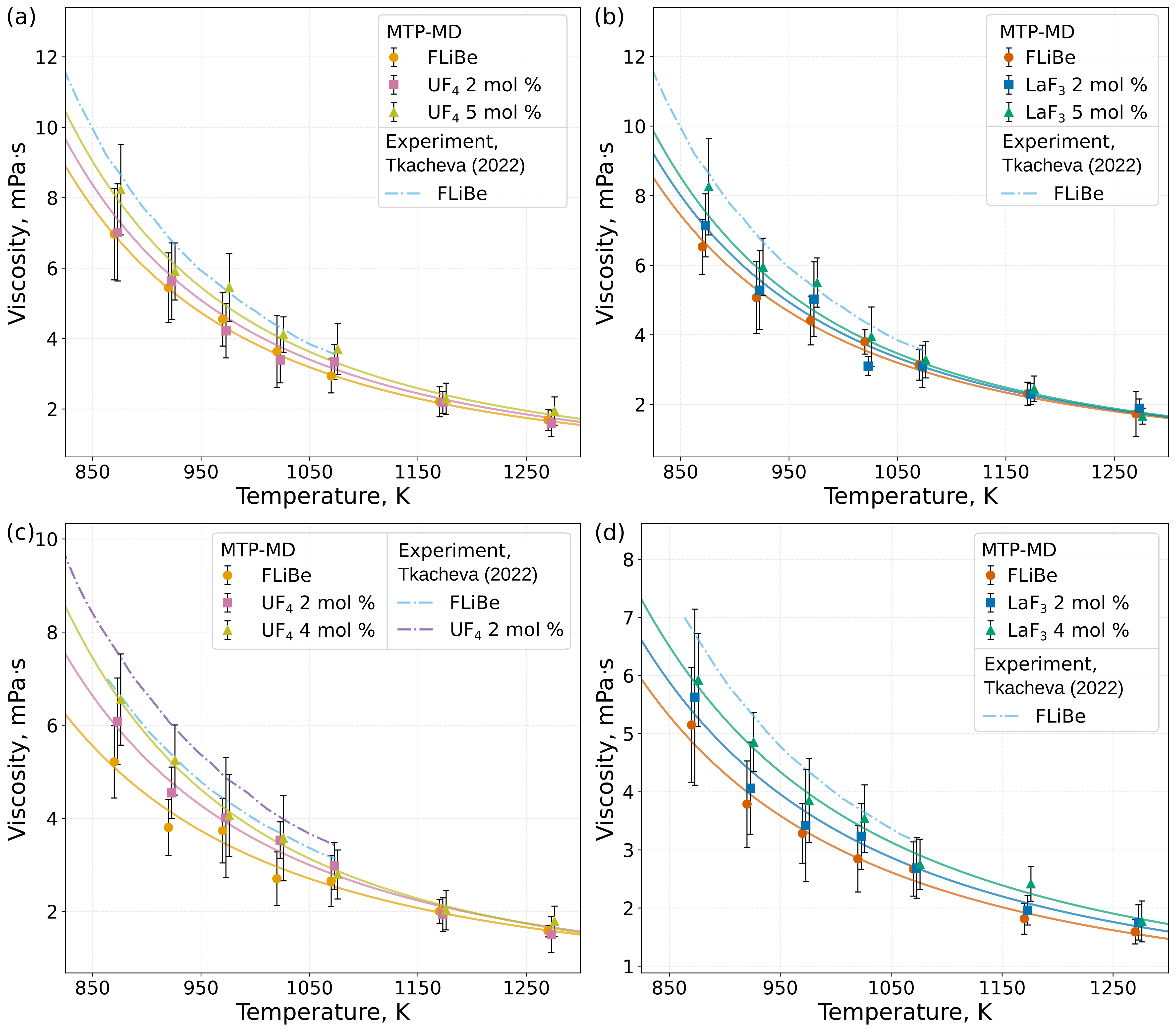}	
	\caption{Temperature dependence of the viscosity of molten salts, obtained with MTP-MD. (a) FLiBe (66~mol~\% -- 34~mol~\%)-UF$_4$, (b) FLiBe (66~mol~\% -- 34~mol~\%)-LaF$_3$, (c) FLiBe (74~mol~\% -- 26~mol~\%)-UF$_4$, and (d) FLiBe (74~mol~\% -- 26~mol~\%)-LaF$_3$. Experimental data for pure FLiBe is taken from Tkacheva (2022) \cite{FLiBe_UF4_viscosity_Tkacheva2022}.} 
	\label{fig:viscosity}
\end{figure*}

We next discuss the dynamic properties, viscosities of FLiBe-LaF$_3$ and FLiBe-UF$_4$ molten salts. The results of calculations are presented in Fig.~\ref{fig:viscosity}. Calculated viscosities for pure FLiBe of both compositions are underestimated by 20-25~\% on average. Notably, the relative error in the calculated viscosity decreases with increasing temperature. However, the physicochemical trends are reproduced accurately. The viscosity decreases with increasing temperature, and the molten salts containing LaF$_3$ and UF$_4$ are more viscous. Experimental measurements \cite{FLiBe_UF4_viscosity_Tkacheva2022} show that, the viscosity of FLiBe 73~mol~\% -- 27~mol~\% increases by 10 \% upon the addition of 2~mol~\% UF$_4$. The MTP-MD predicts similar behavior with the viscosity increase of 13~\% for similar molten salt compositions. The LaF$_3$ additions also increase the viscosity of FLiBe molten salts, but the effect is weaker than that of UF$_4$ addition. The FLiBe 74~mol~\% -- 26~mol~\% viscosity is lower than that of the FLiBe 66~mol~\% -- 34~mol~\%. This can be explained by the lower BeF$_2$ concentration, which, according to the structure analysis and literature data, forms chains and a network in the molten salt. Notably, the viscosity increase upon LaF$_3$ and UF$_4$ addition can also be attributed to the increased number of network-forming atoms in the molten salts.

Viscosity, $\eta$, can be described by Arrhenius-type expression with coefficients that depend on composition:
\begin{equation}
    \eta(T, m) = (\eta_0+\alpha \cdot m) \cdot e^{\frac{E_0+\beta \cdot m}{T}},
    \label{eq:eta_app}
\end{equation}
where $T$ is the temperature, $m$ is the molar fraction of the additive (LaF$_3$ or UF$_4$), $\eta_0$ is the pre-exponential factor for pure FLiBe, $E_0$ is the corresponding activation energy, $\alpha$ and $\beta$ describe the composition-induced shifts of $\eta_0$ and $E_0$. The relationships between viscosity, composition, and temperature established in our work are presented in Tab.~\ref{tab:visc}. 

\begin{table}[h]
\centering
\caption{Coefficients for viscosity calculation according to equations \ref{eq:eta_app} for systems computed in our research. }
\begin{tabular}{|c|cc|cc|}
\hline
Base salt LiF--BeF$_2$            & \multicolumn{2}{c|}{66--34}                & \multicolumn{2}{c|}{74--26}              \\ \hline
Addition             & \multicolumn{1}{c|}{La}           & U            & \multicolumn{1}{c|}{La}          & U            \\ \hline
$\eta_0\cdot 10^{5}$, Pa$\cdot$s & \multicolumn{1}{c|}{9.32}  & 6.99  & \multicolumn{1}{c|}{13.40} & 10.66  \\ \hline
$\alpha\cdot 10^{4}$, Pa$\cdot$s & \multicolumn{1}{c|}{-1.82} & -0.70 & \multicolumn{1}{c|}{4.85} & -10.27\\ \hline
$E_o$, K             & \multicolumn{1}{c|}{3728}         & 4007         & \multicolumn{1}{c|}{3118}        & 3428         \\ \hline
$\beta$, K           & \multicolumn{1}{c|}{3088}         & 4477         & \multicolumn{1}{c|}{1177}        & 16062        \\ \hline
\end{tabular}
\label{tab:visc}
\end{table}

\begin{figure*}
	\centering 
	\includegraphics[trim={0cm 0cm 0cm 0cm}, clip, width=1\linewidth]{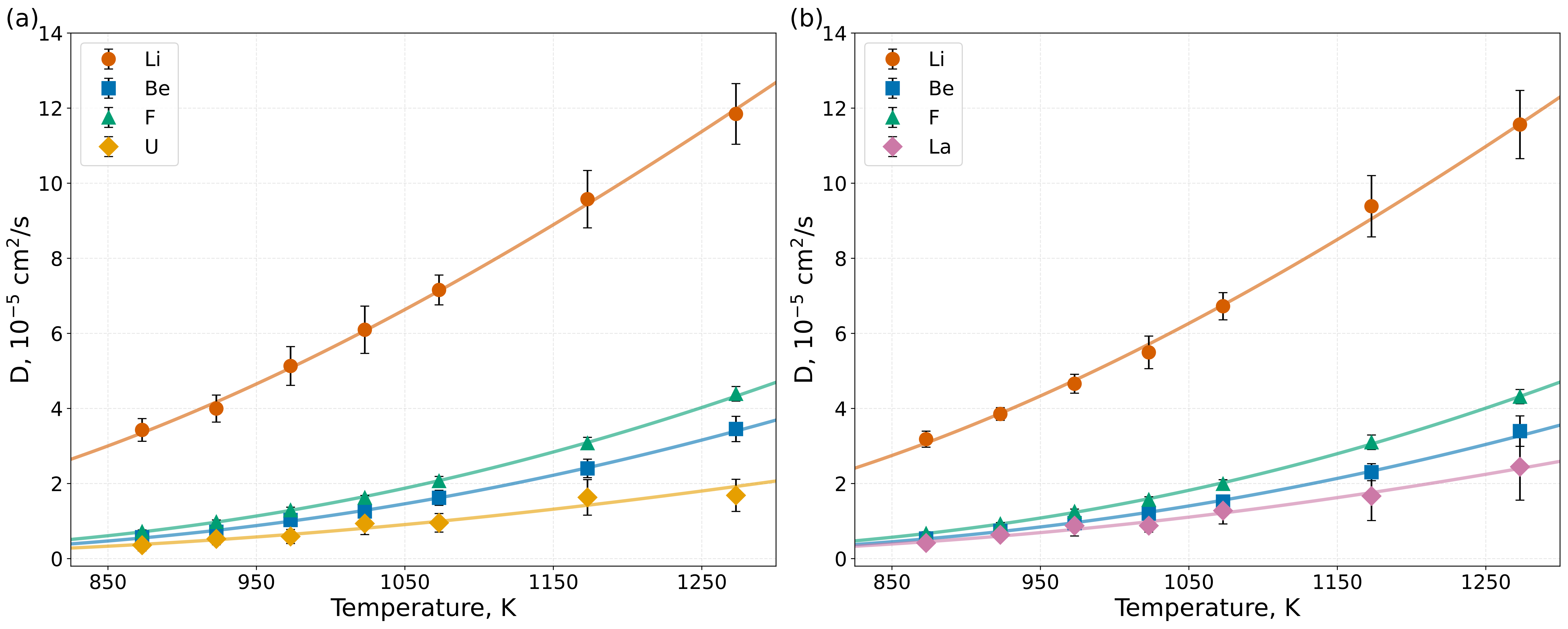}	
	\caption{Temperature dependence of diffusion coefficients of species in FLiBe (66~mol~\% -- 34~mol~\%) with 5~mol~\% additions of (a) UF$_4$ or (b) LaF$_3$.} 
	\label{fig:diffusivity}
\end{figure*}

The transport of mass in the melt is described by the self-diffusion coefficients. Atomic diffusion in a melt is another fundamental property of liquid dynamics that provides important information~\cite{Rollet2009}. The temperature-dependence of self-diffusion coefficients (labeled as D) for FLiBe 66~mol~\% -- 34~mol~\% with 5~mol~\%  additions of UF$_4$ and LaF$_3$ are reported in Fig.~\ref{fig:diffusivity}, the corresponding data for FLiBe 74~mol~\%--26~mol~\% composition is provided in the SI. According to our calculations, the diffusion coefficients obey the Arrhenius law, the approximating coefficients are reported in the SI. The diffusion coefficients of species are ordered as follows $\text{D}_{\text{Li}}>\text{D}_{\text{F}}>\text{D}_{\text{Be}}>\text{D}_{\text{U}}\sim\text{D}_{\text{La}}$. $\text{D}_{\text{Li}}$ is $\sim$~4 times higher than $\text{D}_{\text{Be}}$ and $\text{D}_{\text{F}}$; this is in an agreement with Be atoms forming clusters and a network structure in the melt, slowing the diffusion. The same behavior is observed in an experimental study \cite{mei2013investigation}. However, because the viscosity is underestimated, the calculated diffusion coefficients are correspondingly higher. The same overestimation was observed in previous computational study of FLiBe \cite{attarian2022thermophysical}. The La and U atoms have similar diffusion coefficients to each other, and both are even less mobile than Be atoms. The large and heavy clusters formed by these elements in the molten salt, discussed above, suppress diffusion. Notably, the diffusion coefficients for the FLiBe 74~mol~\% -- 26~mol~\% are higher for all atomic species, that is consistent with the lower viscosity of molten salts.

\begin{figure*}
	\centering 
	\includegraphics[width=1\linewidth]{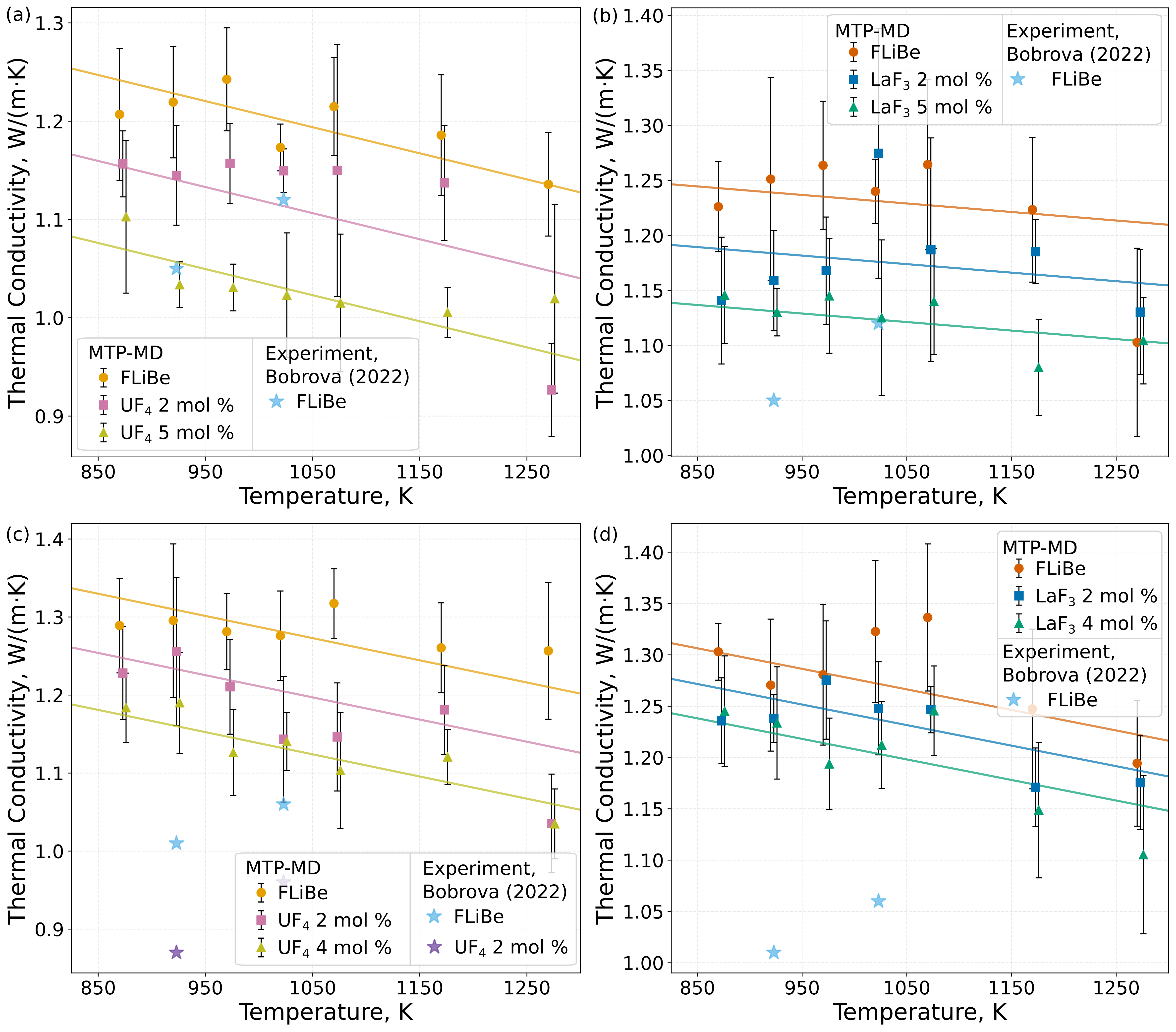}	
	\caption{Temperature dependence of the thermal conductivity of molten salts, obtained with MTP-MD. (a) FLiBe (66~mol~\% -- 34~mol~\%)-UF$_4$, (b) FLiBe (66~mol~\% -- 34~mol~\%)-LaF$_3$, (c) FLiBe (74~mol~\% -- 26~mol~\%)-UF$_4$, and (d) FLiBe (74~mol~\% -- 26~mol~\%)-LaF$_3$. Experimental data for pure FLiBe is taken from Bobrova (2022) \cite{bobrova2023thermophysical}.} 
	\label{fig:thermal_conductivity}
\end{figure*}

We now present the thermal conductivity, the most experimentally debated of the transport properties.  Fig.~\ref{fig:thermal_conductivity} shows our calculated results (the linear fit parameters are listed in Tab.~\ref{tab:TC}). We observe lower thermal conductivity at higher temperatures for the molten salts. The temperature dependence of thermal conductivity in molten salts remains a subject of debate in the literature. In the experimental studies \cite{bobrova2023thermophysical, rudenko2024density, kesler2025radiative} a positive temperature dependence was reported, whereas a constant value of 1~W/(m$\cdot$K)  is recommended in Ref.~\cite{williams2006assessment}, while a negative slope of thermal conductivity versus temperature is observed in both computational, theoretical and experimental studies \cite{zakiryanov2024structure, denisov2025thermal, pan2021finite,YANG2023158}. Regarding the LaF$_3$ and UF$_4$ additions, both additives decrease the thermal conductivity of molten salts. This result is in agreement with both experimental and computational studies on molten salts \cite{bobrova2023thermophysical, kesler2025radiative, denisov2025thermal}. This behavior can be attributed to the U-F and La-F complexes discussed above. The formation of large, heavy La-F and U-F complexes disrupts thermal transport in FLiBe by acting as massive phonon-like scattering centers. These centers sever the continuous Be-F-Be bridging network and localize vibrational energy, thereby drastically reducing the phonon mean free path \cite{denisov2025thermal}. Despite the good qualitative agreement, we observe a non-negligible relative error in the predicted thermal conductivity. For all the simulated state points the thermal conductivity is overestimated by 20-30~\%. Notably, the results of experimental studies can differ among each other by up to 50~\% \cite{YANG2023158}.

Thermal conductivity is described using a linear equation:  
\begin{equation}
    \kappa(T, m) = (\kappa_0+\alpha \cdot m) + \beta\cdot T,
    \label{eq:TC_app}
\end{equation}
where $T$ is the temperature, $m$ is the molar fraction of the additive (LaF$_3$ or UF$_4$), $\kappa_0$ is the empirical zero-temperature intercept (a fitting parameter with no physical meaning at actual 0~K, since FLiBe is not liquid at that temperature), $\alpha$ accounts for the effect of composition, and $\beta$ is the common temperature coefficient, which is assumed to be independent of $m$ to avoid overparameterization.

\begin{table}[h]
\centering
\caption{Coefficients for thermal conductivity calculation according to equations \ref{eq:TC_app} for systems computed in our research. }
\begin{tabular}{|c|cc|cc|}
\hline
Base salt  LiF--BeF$_2$                   & \multicolumn{2}{c|}{66--34}                & \multicolumn{2}{c|}{74--26}               \\ \hline
Addition                    & \multicolumn{1}{c|}{La}           & U            & \multicolumn{1}{c|}{La}           & U            \\ \hline
$\kappa_0$, W/(m$\cdot$K)  & \multicolumn{1}{c|}{1.42}       & 1.54       & \multicolumn{1}{c|}{1.67}       & 1.58       \\ \hline
$\alpha$, W/(m$\cdot$K)     & \multicolumn{1}{c|}{-2.22}      & -3.06      & \multicolumn{1}{c|}{-2.21}      & -4.16      \\ \hline
$\beta\cdot 10^{4}$, W/(m$\cdot$K$^2$) & \multicolumn{1}{c|}{-1.8} & -3.3 & \multicolumn{1}{c|}{-3.7} & -2.9 \\ \hline
\end{tabular}
\label{tab:TC}
\end{table}

Overall, the physicochemical properties of FLiBe 66~mol~\% -- 34~mol~\% and 74~mol~\% -- 26~mol~\% molten salts with additions of LaF$_3$ or UF$_4$ derived from MTP-MD simulations are reported in terms of coefficients that approximate the temperature (in the range of 870-1300~K) and composition (up to 5~mol~\% of LaF$_3$ or UF$_4$) dependencies. The coefficients for density, viscosity, and thermal conductivity for the systems with UF$_4$ and LaF$_3$ are summarized in Tab.~\ref{tab:dens}, Tab.~\ref{tab:visc}, and Tab.~\ref{tab:TC}, respectively. These equations are used in FE modeling.

\subsection{Validation of the finite-elements model}

For the validation against experiment, the transport properties of FLiBe are taken from the literature (denoted \emph{FE-ref}) (see Supplementary Materials). The corresponding simulation in which the same fluid is described by MTP-MD-derived properties is denoted \emph{FE-MD} and is reported in Section ~\ref{AMR} for completeness.

\begin{figure}[h]
  \centering
  \includegraphics[width=1\columnwidth]{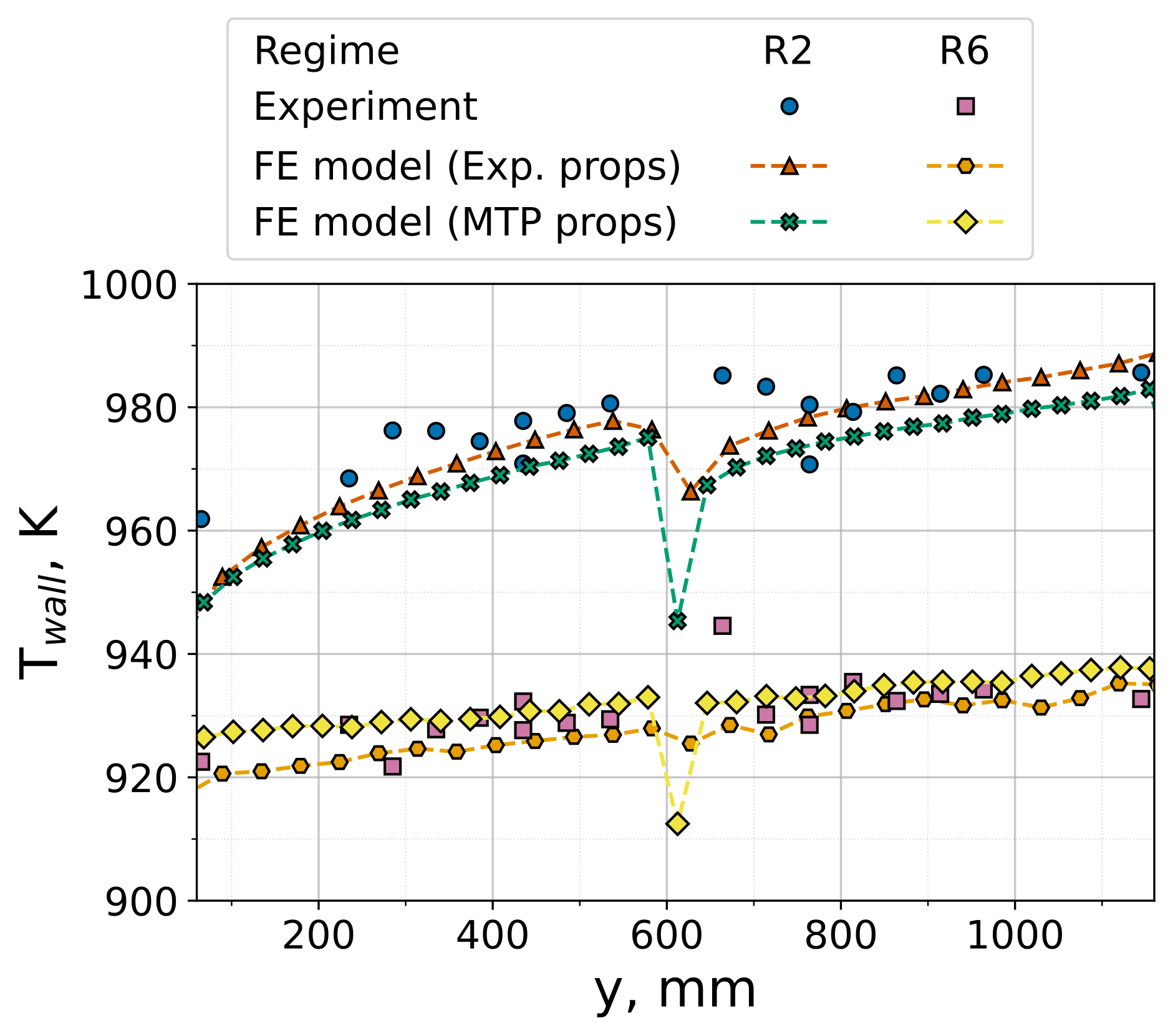}
  \caption{Wall temperature along the pipe length for regimes R2 (500<$Re$<1300) and R6 (7000<$Re$<11000), see Table~\ref{tab:regimes} for operating conditions. Experimental data: blue circles (R2) and purple squares (R6). FE predictions using experimental properties: vermilion triangles (R2) and golden orange hexagons (R6). FE predictions using MTP-MD-derived properties: green filled crosses (R2) and yellow diamonds (R6).}
  \label{fig:validation_Twalvsyl}
\end{figure}

The validation proceeds in three steps: outer-wall temperature, local heat-transfer coefficient, and the integrated $Nu$--$Re$ criterion correlation, each compared with the corresponding experimental measurement. We started coupling results of the atomistic simulations with the finite-element model by performing a validation simulation for reference parameter -- pipe wall temperature. We calculated the temperature of the outer pipe wall and compared the results with measured thermocouple data. The pipe wall temperature, measured experimentally and obtained from the model for regimes with $Re > 500$ (laminar) and $Re > 7000$ (turbulent). is shown in Fig. \ref{fig:validation_Twalvsyl}. The figure also shows similar characteristics obtained for \emph{FE-MD} model.

In both cases, the numerical model successfully reproduces the experimentally observed trends. The wall temperature increases gradually along the flow direction as the molten salt absorbs heat from the heated wall.

To further validate the approach on the experimental data, which was obtained for pure FLiBe, we compared the thermophysical parameters of the developed model with experimental data. The local heat-transfer coefficient $\alpha_{\rm loc}$ and the dimensionless numbers $Nu$, $Re$, are defined in Section~\ref{subsubsec:fe_metrics} (eq.~\ref{eq:alpha_loc} and eq.~\ref{eq:dimensionless}).   

\begin{figure}[h] 
  \centering
  \includegraphics[width=1\columnwidth]{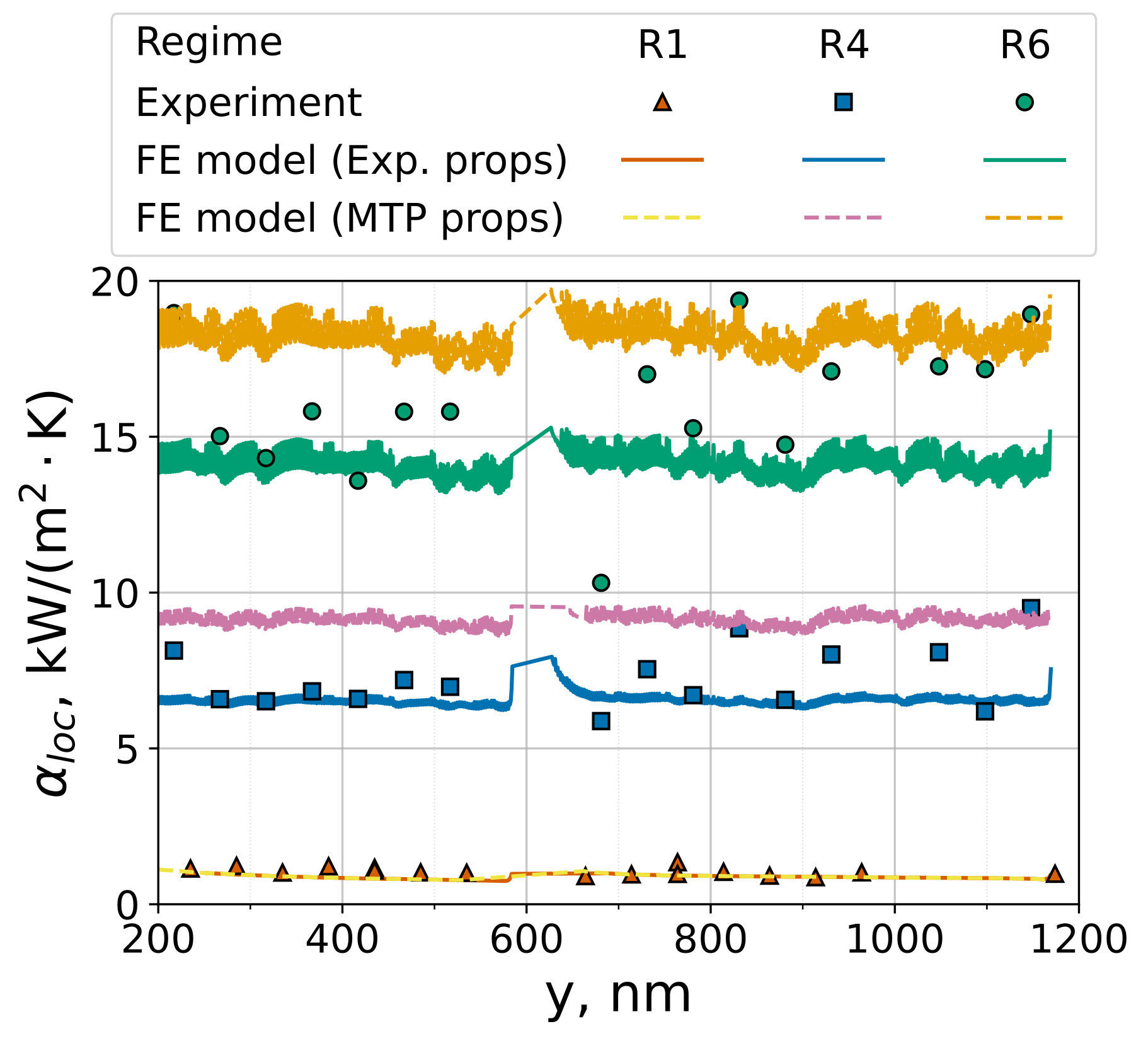} 
  \caption{Local heat transfer coefficient $\alpha_{loc}$ along the pipe length for regimes R1 (500<$Re$<1300), R4 (3300<$Re$<6000), and R6 (7000<$Re$<11000), operating conditions provided in Tab.~\ref{tab:regimes}. Experimental data are shown by filled markers: green circles (R1), blue triangles (R4), and vermilion squares (R6). Finite-element model predictions using experimental thermophysical properties are shown by solid lines in the corresponding experimental colors (green (R1), blue (R4), and vermilion (R6)). FE predictions using MTP-MD-derived properties are shown by dashed lines in alternative colors: yellow (R1), purple (R4), and golden orange (R6).}
  \label{fig:alpha_local}
\end{figure}

Fig.~\ref{fig:alpha_local} presents local values of the heat transfer coefficients calculated in the model and obtained from the thermal test facility in the experiment for regimes with $500 < Re < 1300$, $3300 < Re < 6000$ and $7000 < Re < 11000$. The peaks observed in the central part of the dependencies are attributable to the presence of a steel current collector in this region, resulting in localized heat flux intensification. Overall, the \emph{FE-ref} model shows good agreement with the experimental data across the entire investigated operating range. The most significant deviations occur at the highest flow rates, where uncertainty reaches 18~\% (relative to the average experimental data). This discrepancy stems from several factors; most notably, the experimental data exhibit increasing scatter at higher mass flow rates. Increased linear velocity at these rates prevents the thermal regime from reaching a steady state, thereby increasing measurement dispersion. Conversely, the model is solved as a steady-state problem with prescribed boundary conditions, and therefore produces a smooth, deterministic profile of local heat transfer values along the pipe length without the statistical scatter of the experiments. Furthermore, discrepancies may arise from the idealized boundary conditions used in the simulations and subtle variations in actual flow conditions that are difficult to reproduce numerically. Nevertheless, the finite-element model accurately captures both the qualitative trends and quantitative values of the measured parameters, validating its suitability for future heat transfer simulations of FLiBe-based molten salts with modified compositions. 

The figure also shows the heat exchange coefficients for \emph{FE-MD} model obtained with the MTP-MD results for pure FliBe (66~mol~\% -- 34~mol~\%). Comparison with FE-simulation data reveals a systematic overestimation of values by an average of 25-28~\%, with the magnitude of the error independent of the studied regime. As discussed in section (Section~\ref{AMR}), when determining the melt transport properties, we can expect an uncertainty of up to 30~\%. In particular, overestimated thermal conductivity values lead to increased heat transfer intensity between the bulk and the wall layer of the coolant.

\begin{figure}[h] 
  \centering
  \includegraphics[width=1\columnwidth]{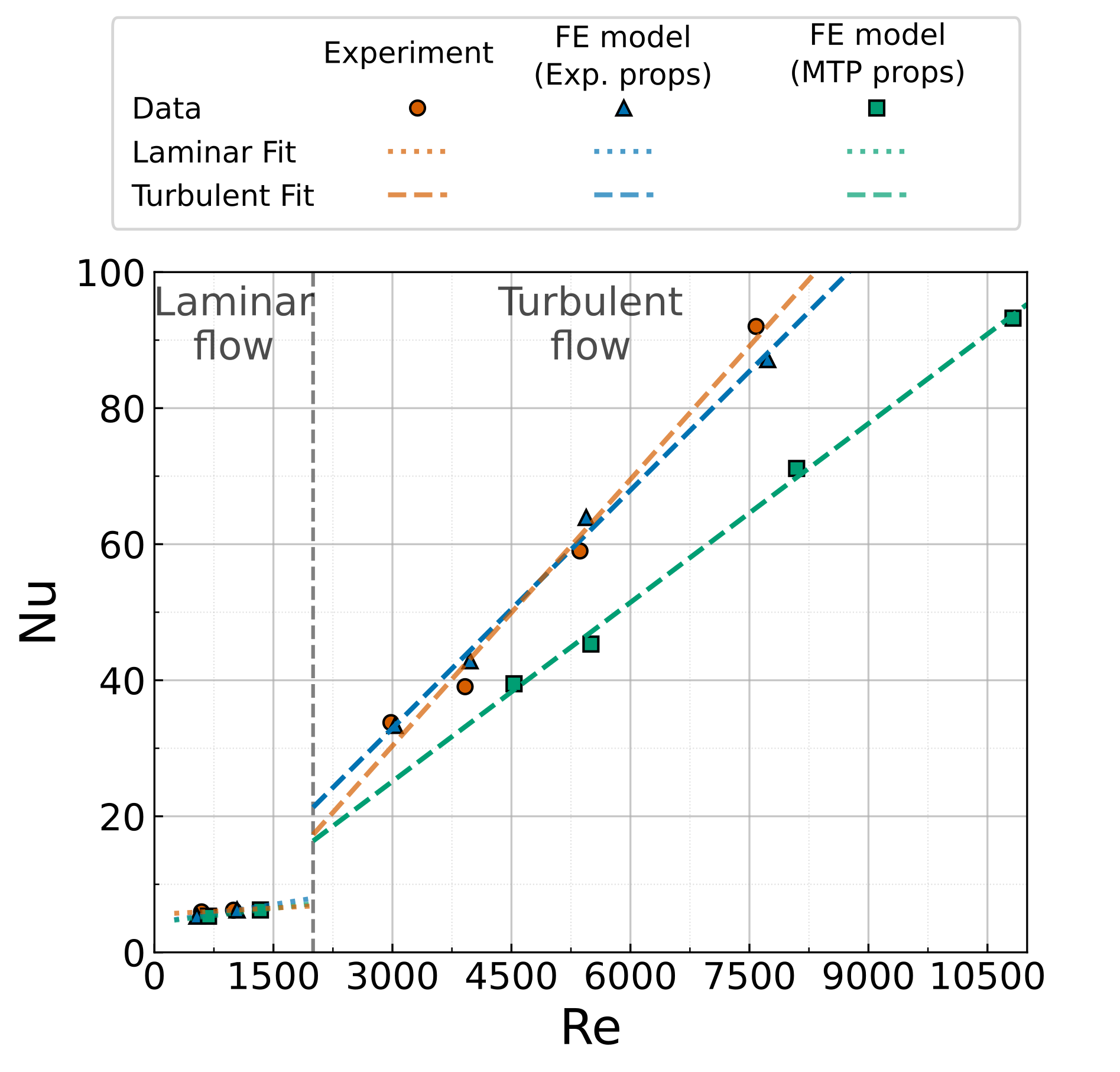} 
  \caption{Nusselt number ($Nu$) as a function of Reynolds number ($Re$) for all six regimes (see Table~\ref{tab:regimes} for operating conditions). Data points represent values averaged along the pipe length, obtained from experiments (vermilion circles), FE modeling using experimental thermophysical properties (blue triangles), and FE modeling using MTP-derived properties (green squares). The dotted and dashed lines correspond to linear fits for laminar ($Re<2300$) and turbulent ($Re>2300$) regimes, respectively.} 
  \label{fig:Nu-Re}
\end{figure}

The Fig.~\ref{fig:Nu-Re}  shows the criterion dependencies $Nu(Re)$ derived using the Nusselt and Reynolds number values averaged along the pipe length for all regimes under investigation  
To jointly assess the contribution of uncertainty in the heat transfer fluid properties to the convective heat transfer parameters, the figure presents the results of the MTP-MD modeling. Similar to the local heat transfer coefficient, the predicted values of the criterion dependencies  are systematically overestimated, exhibiting a shift upward and to the right relative to the experimental and modeling data. The Reynolds number is overestimated by an average of 28~\% (rightward shift), regardless of the flow regime. In this case, the dominant source of uncertainty is the determination of the melt viscosity.
The FE modeling results can be interpreted as follows: the uncertainty observed in both the local heat transfer parameters and the averaged dimensionless characteristics is in good agreement with the uncertainty associated with calculating the heat transfer fluid properties and is independent of the flow regime.
Nevertheless, the qualitative behavior predicted by the thermophysical model appears physically reasonable. Therefore, the next step is to evaluate the influence of the LaF$_3$ and UF$_4$ additives on the heat transfer performance of the test rig.

\subsection{Results  of the finite-elements modeling with input from atomistic simulations}
\label{sec:fe-md}

We used results of MTP-MD to obtain the heat exchange coefficients $\alpha_{loc}$ and criterion correlations $Nu(Re)$ when FLiBe-UF$_4$ (0, $\sim$2, $\sim$5~mol~\%) and FLiBe-LaF$_3$ (0, $\sim$2, $\sim$5~mol~\%) are coolants in the digital twin model of the thermal test facility.  

\begin{figure}[h] 
  \centering
  \includegraphics[width=1\columnwidth]{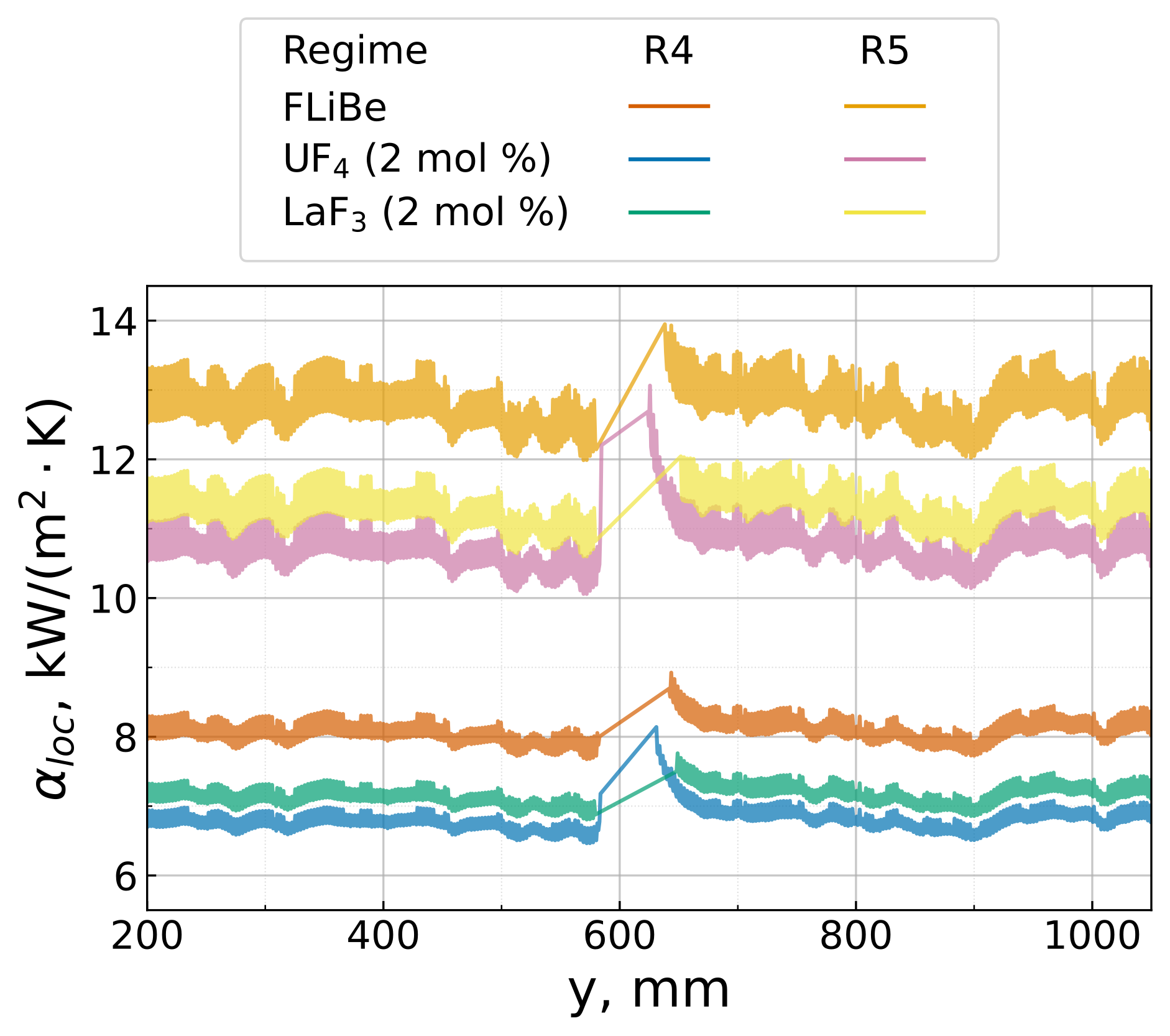} 
  \caption{Local heat transfer coefficient $\alpha_{loc}$ along the pipe length for regimes R4 (3300<$Re$<6000) and R5 (5000<$Re$<8000), operating conditions provided in Tab.~\ref{tab:regimes}, computed for various coolants. Finite-element model predictions using MTP-MD-derived thermophysical properties are shown by solid lines for pure FLiBe 66~mol~\% -- 34~mol~\%, FLiBe - 2~mol~\% LaF$_3$, and FLiBe - 2~mol~\% UF$_4$.}
  \label{fig:alpha_coeff}
\end{figure}

Fig.~\ref{fig:alpha_coeff} shows the calculated local heat transfer coefficients $\alpha_{loc}$ for FLiBe-UF$_4$ (2~mol~\%) and FLiBe-LaF$_3$ (2~mol~\%) mol\-ten systems. Regimes correspond to the Reynolds numbers $Re \approx 4500$ and $Re \approx 7500$, respectively. As the flow turbulizes, the heat transfer coefficient increases by about 40~\%, which is consistent with the behavior of the system using the base FLiBe melt (see Fig.~\ref{fig:alpha_local}). 
In terms of heat transfer efficiency reduction, the systems under study can be arranged as follows: $\alpha_{\text{FLiBe}} > \alpha_{\text{FLiBe-LaF}_3} > \alpha_{\text{FLiBe-UF}_4}$. Consequently, under the same coolant flow regimes, varying the coolant composition within the solubility limits of the components yields a reduction in heat transfer efficiency of  approximately 8\% (with the addition of 2~mol~\% LaF$_3$) and 10\% (with the addition of 2~mol~\% UF$_4$). The addition of 5~mol~\% of LaF$_3$ and UF$_4$ reduces heat transfer efficiency by 10\% and 11\%, respectively.

In the classical heat transfer study \cite{Cooke1973ForcedconvectionHM}, key relationships connecting the Nusselt number with the Rey\-nolds and Prandtl numbers were established. The experimental data generalized in that work demonstrated that in laminar ($Re$<2300) and turbulent (2300<$Re$<11000) regimes, the behavior of the molten salt is well described by standard correlations (eqs.~\ref{eq:sieder-tate}, \ref{eq:hausen}). However, in the transition region (2300<$Re$<4000), a systematic delay in the development of turbulence was identified, attributed to the stabilization of the laminar boundary layer due to viscosity variations.

Building on these findings, the present work compared the heat transfer data for the studied systems within the Reynolds number range from 0 to 11000 with the correlations from \cite{Cooke1973ForcedconvectionHM}, including Sieder-Tate equation  \cite{sieder1936heat} for low Reynolds number and Hausen equation for high Reynolds number \cite{hausen1959neue}  and assessed their validity under the conditions of our experiment.

Two empirical correlations are used to interpret the Nusselt number $Nu_{theor}$. The Sieder--Tate correlation eq.~\ref{eq:sieder-tate} is valid for $Re < 2300$ (laminar); the Hausen correlation eq.~\ref{eq:hausen} is valid for $2300 < Re < 11000$ (transitional and turbulent):

\begin{equation}
Nu_{theor} = 1.86 \left[ Re \cdot Pr \left( \frac{D}{L} \right) \right]^{1/3} \left( \frac{\mu}{\mu_s} \right)^{0.14}
         \label{eq:sieder-tate}
\end{equation}

\begin{equation}
Nu_{theor} = 0.116 \left( Re^{2/3} - 125 \right) Pr^{1/3} \left( \frac{\mu}{\mu_s} \right)^{0.14}
     \label{eq:hausen}
\end{equation}

\noindent where \(D\) is the pipe diameter, \(L\) is the pipe length, \(\mu\) is the dynamic viscosity at bulk temperature, and \(\mu_s\) is the dynamic viscosity at wall temperature.

The dependence of the heat transfer function (HTF) on the Reynolds number is calculated according to:

\begin{equation}
\text{HTF} = \frac{Nu_{theor}}{\left(Pr\right)^{1/3} \left(\frac{\mu}{\mu_s}\right)^{0.14}}.
\label{eq:HTF}
\end{equation}

\begin{figure}[h] 
  \centering
  \includegraphics[width=1\columnwidth]{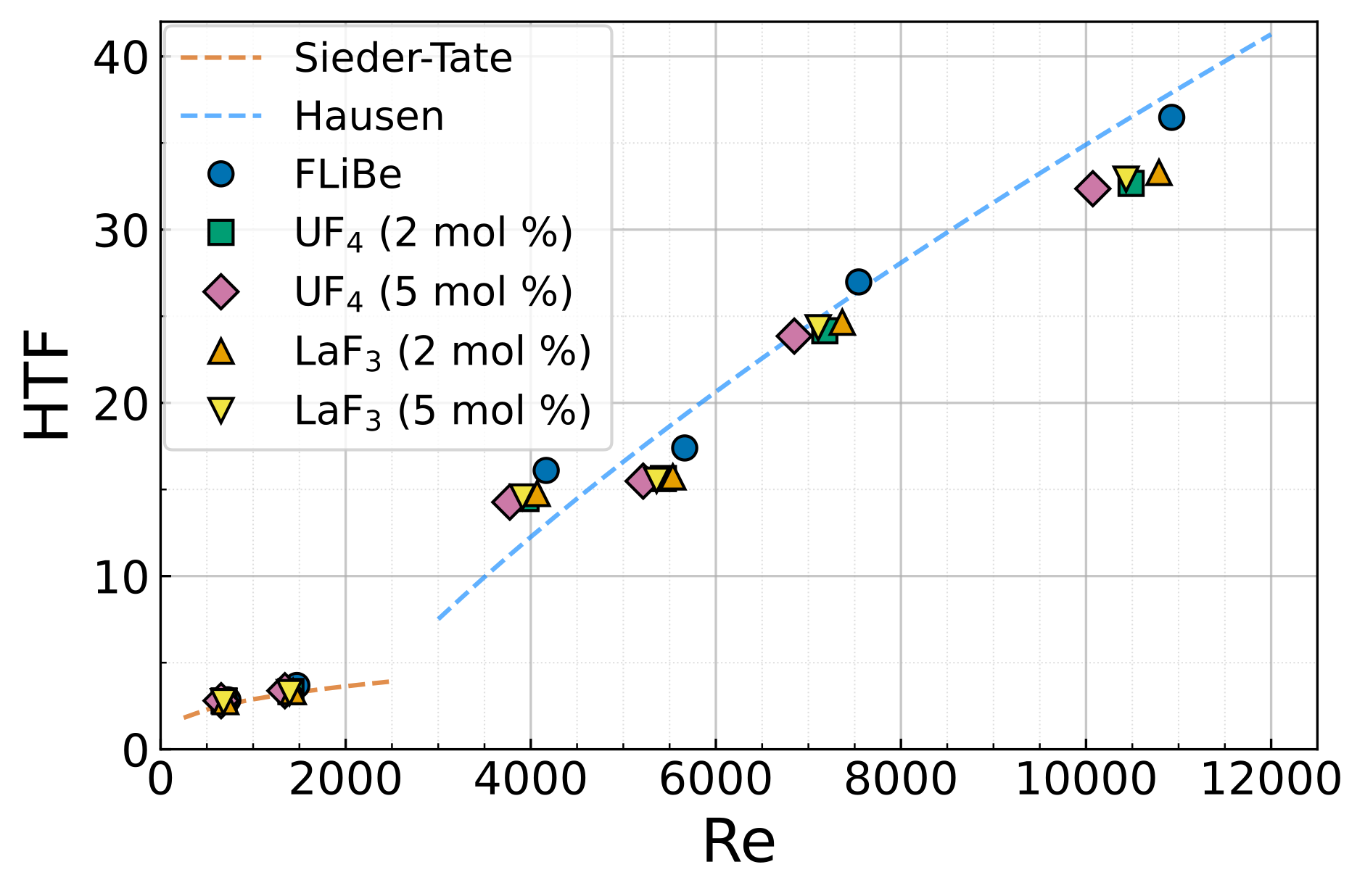} 
  \caption{Heat transfer functions for the coolants: FLiBe and FLiBe containing 2 and 5~mol~\% of LaF$_3$ or UF$_4$, calculated in the model (points) and estimated from the theory (dotted lines).} 
  \label{fig:HTF_final}
\end{figure}

Figure \ref{fig:HTF_final} presents our calculated results together with data estimated according to eqs.~\ref{eq:sieder-tate}-\ref{eq:hausen}. The comparison results showed that in the laminar regimes, our data are in good agreement with eq. \ref{eq:sieder-tate}; however, in the transitional regimes, a systematic deviation from eq. \ref{eq:hausen} is observed. The boundary layer development problem is most pronounced in the Reynolds number range from 2300 to 4000, where entrance effects persist over the entire length of the thermal test section. The same effect may be observed up to \(Re = 5000\) at higher heat fluxes.

Our data lie above the standard correlations in the transition region but not in the laminar region. If the deviations in our data were caused by heat conducting surface films or entrapped gas, one would expect this effect to manifest across all flow regions. In the laminar region, the heat transfer function is virtually unaffected by viscosity. In contrast, under transitional and turbulent flow conditions, this dependence is pronounced. Consequently, the deviation between the HTF performance calculated from our model and that evaluated via eq.~\ref{eq:HTF} could arise from uncertainties in the salt viscosity data at the relevant operating temperatures. Furthermore, existing theoretical models do not fully describe the dependence in the transition regime, indicating the need for their further development and refinement.

The transitional flow regime of \(Re = 2300\)-\(5000\) is highly sensitive to the geometric parameters of the working section, which determine the magnitude of the heat flux density across the heat transfer boundary. In turn, the heat flux intensity determines the development of a viscous sublayer near the wall, where heat transfer is determined by the thermal conductivity of the fluid, and the flow regime is laminar. In the transient flow regime, the heat transfer intensity in the radial direction increases, meaning the bulk temperature of the coolant decreases but the velocities are still insufficient; the viscous sublayer is sufficiently developed and can significantly reduce heat transfer. However, the viscous sublayer can be unstable and exhibit pulsations, which complicates the shape of the heat transfer function and increases uncertainty when operating in the transient flow ranges.

\section{Conclusions}

We have developed a hierarchically integrated multiscale framework that couples machine-learning-driven atomistic simulations to continuum finite-element modeling for heat-transfer prediction in FLiBe-based molten salts. At the atomistic level, MTPs were trained for pure FLiBe (66~mol~\% -- 34~mol~\% and 74~mol~\% -- 26~mol~\% LiF-BeF$_2$ ratios), FLiBe-LaF$_3$, and FLiBe-UF$_4$ melts and used to compute temperature- and com\-po\-si\-ti\-on-dependent transport properties (density, viscosity, thermal conductivity, and isobaric heat capacity). These properties were mapped as inputs into a three-dimensional FE model of an experimental thermal loop.

The \emph{FE-ref} configuration, which uses reference literature transport properties for pure FLiBe 66~mol~\% -- 34~mol~\%, reproduces the experimental heat-transfer behavior to within 10~\% in laminar regime and up to 18~\% in the transitional and turbulent regimes. The end-to-end pipeline is therefore validated for the 66~mol~\% -- 34~mol~\% composition. The 74~mol~\% -- 26~mol~\% composition is supported at the atomistic level only; its FE validation is left for future work. The \emph{FE-MD} configuration, which uses the MTP-MD-derived transport properties of Section~\ref{AMR}, systematically overestimates the heat-transfer coefficient by 25--28~\% and shifts the $Nu$--$Re$ curve rightward by 28~\%. These offsets are consistent with the MTP-MD biases on thermal conductivity ($+$20--30~\% relative to experiment) and viscosity ($-$20--25~\%), which compound in the same direction to increase the predicted $\alpha_{\rm loc}$. The qualitative ordering $\alpha_{\rm FLiBe} > \alpha_{\rm FLiBe\text{--}LaF_3} > \alpha_{\rm FLiBe\text{--}UF_4}$ is the more robust conclusion from the MTP-MD side; the absolute value of any single-component $\alpha$ is sensitive to the MTP accuracy.

Applied to the ternary systems FLiBe-LaF$_3$ and FLiBe-UF$_4$ over the investigated compositional window (0-5~mol~\%), the FE-MD model predicts a mean reduction in heat-transfer efficiency of 8-11~\% relative to pure FLiBe. The degradation is driven primarily by increased viscosity and decreased thermal conductivity, with UF$_4$-bearing melts exhibiting the most pronounced effect. The 8-11~\% range should be read as the model's prediction given the MTP-MD biases, not as an experimental fact.

A limitation of the presented end-to-end approach for quantitatively assessing heat transfer parameters is the treatment of regimes with $Re$ > 2300, where the FE model deviates from the Sieder--Tate and Hausen correlations. The deviation is attributable to two effects: the MTP-MD biases on the salt transport properties, and the approximate treatment of boundary-layer transition by the $k$--$\epsilon$ model. The fact that the FE-ref configuration (which uses reference properties) also shows a deviation in the transition regime indicates that the second effect is the dominant one. Refining the turbulence model and improving the MTP accuracy are the two natural next steps.

\section{Data Availability}

The MTPs and FE model files generated in this study are available from the corresponding author upon reasonable request.

\section*{Author Contributions}
\label{sec:contributions}

Mikhail Polovinkin: methodology, investigation (developed the MTPs, conducted and processed MD modeling), validation, conceptualization, data curation, writing -- original draft. Ksenia Abramova: methodology, investigation (developed the FE model), validation, conceptualization, data curation, writing -- original draft. Oksana Rahmanova: methodology, investigation, validation, conceptualization. Farit Valiev: investigation, writing -- original draft. Andrey Isakov: conceptualization. Andrey Goryachikh: conceptualization. Alexander Galashev: funding acquisition, validation, conceptualization. Yurii Zaikov: conceptualization. Dmitrii Maksimov: conceptualization, writing -- original draft. Alexander Shapeev: resources, supervision. Nikita Rybin: conceptualization, supervision, writing -- review and editing.

\section*{Acknowledgments}
Ksenia Abramova, Oksana Rahmanova, Farit Valiev, and Alexander Galashev acknowledge financial support of the Russian Science Foundation, project No. 25-23-01312.

\bibliographystyle{elsarticle-harv} 
\clearpage
\bibliography{example}

\clearpage

\onecolumn
\appendix

\section{Finite-element model: governing equations, material properties, and boundary conditions}
\label{SI:femodel}

This section collects, in one place, the governing equations, the material properties, and the boundary conditions used in the FE simulations. It is the supporting information for Section~2 of the main text. The numerical implementation (mesh strategy, turbulence model, and solver settings) is described in the next sections.

\subsection*{Governing equations}

The steady-state FE problem couples mass conservation, momentum conservation (Navier--Stokes), energy conservation (with Joule heating), and the electric current conservation equation. 

The numerical model was developed in COMSOL Multiphysics using a coupled multiphysics approach. The simulation combined the Fluid flow Module (Turbulent Flow, $k-\epsilon$ model), Heat Transfer Module, AC/DC Module, and the predefined multiphysics interfaces Electromagnetic Heating and Non-Isothermal Flow.

The steady-state current-conservation equation is:

\begin{equation}
\nabla\cdot\mathbf{J} = 0,
\label{eq:SI-continuity}
\end{equation}

\noindent where $\mathbf{J}$ is the current density vector (A/m$^2$).

The steady-state mass-conservation (continuity) equation is:

\begin{equation}
\nabla\cdot(\rho\,\mathbf{u}) = 0,
\label{eq:SI-continuity}
\end{equation}

\noindent where $\mathbf{u}$ is the velocity vector (m/s) and $\rho$ is the salt density (kg/m$^3$).

The fluid flow was modeled as incompressible turbulent flow by solving the Reynolds-averaged Navier–Stokes (RANS) equations together with the $k-\epsilon$ turbulence model:

\begin{equation}
\rho\,(\mathbf{u}\cdot\nabla)\mathbf{u} = -\nabla p + \nabla\cdot\!\left[(\mu\!+\mu_{t}\left)(\nabla\mathbf{u}+\nabla\mathbf{u}^{\mathrm{T}}\right)\right],
\label{eq:SI-momentum}
\end{equation}

where ($\mathbf{u}$) is the velocity vector, (p) is the pressure, ($\rho$) is the density, ($\mu$) is the dynamic viscosity, and ($\mu_t$) is the turbulent viscosity determined from the transport equations for the turbulent kinetic energy (k) and the specific dissipation rate ($\epsilon$).

For the laminar-flow regimes (Re < 2300), the system reduces to the steady incompressible Navier--Stokes equations with no turbulence model.

The steady-state energy-conservation equation with Joule heating is:

\begin{equation}
\rho C_p\,(\mathbf{u}\cdot\nabla)T = \nabla\cdot\!\left[k\,\nabla T\right] + Q_J,
\label{eq:SI-energy}
\end{equation}

\noindent where $C_p$ is the isobaric heat capacity of the salt (J/(kg·K)), $k$ is the thermal conductivity (W/(m·K)), and $Q_J$ is the volumetric Joule-heat source (W/m$^3$).

\subsection*{Material properties}

The equations of the thermophysical problem are solved in the domains of the molten salt, the steel heat-exchanger wall, and the kaolin-wool insulation. The solid materials for current supply (AISI 321) and insulation (50Al\textsubscript{2}O\textsubscript{3}--50SiO\textsubscript{2}) are taken from the COMSOL material library at the relevant temperatures, properties for AISI 316L pipeline are given in Table\ref{tab:AISI_k}, \ref{tab:AISI_sigma}. The temperature-dependent salt properties for \emph{FE-ref}-model (validation model) are given in Table \ref{tab:properties}. The temperature-dependent salt properties for \emph{FE-MTP}-model (coolant with MTP-MD calculated properties) are given in Section 3.1 the main text. 

\subsection*{Boundary and operating conditions}

The boundary conditions applied in the FE simulations are summarised below.

\paragraph{Melt inlet (right end of the working section).} A uniform velocity profile corresponding to the experimental mass flow rate $Q_0$ (Table~3 of the main text) is imposed, together with the experimental inlet temperature $T_{\rm in}$. The direction of the velocity is normal to the inlet cross-section. 

\paragraph{Melt outlet (left end of the working section)} An open-boundary condition (constant static pressure) is applied at the outlet. The thermodynamic state at the outlet is the Open-Boundary condition.

\paragraph{No-slip condition on the inner wall of the steel pipe} Applied in both laminar and turbulent regimes on the inner wall of the steel pipe.

\paragraph{Electrical boundary conditions} Prescribed voltage drop $V_o$ at the two end current supplies, electric ground $V = 0$ at the central current supply, and electrical insulation on all external surfaces. The prescribed voltage drop $V_o$ and the mass flow rate $Q_o$ for each of the six investigated regimes are given in Table~3 of the main text.

\paragraph{External surfaces of the insulation} Convective heat-flux boundary condition (Eq.~(2) of the main text) with $T_{\rm amb} = 300$~K. The natural-convection coefficient $\alpha$ on the outer surface is taken from standard correlations for the orientation of the insulation (horizontal cylinder); the exact value is reported in the COMSOL setup. 


The 250~W heat-loss calibration is enforced via an Open Boundary condition at a specified outlet temperature $T_{upstr}$ = T($R_{therm}$), $R_{therm}$ is the average thermal resistance of the experimental setup)

The average value of the thermal resistance over all experimental regimes is 0.0051 K/W, the $T_{upstr}$ defined as follows: 

\begin{equation}
   T_{upstr} = T_{in} + \left\langle Q_{loss, exp} \right\rangle \cdot \left\langle R_{therm, exp} \right\rangle,
    \label{eq:14}
\end{equation}

The values of the functional dependence (\ref{eq:14}) are determined by the design of the test facility and do not depend on the flow regime.

\subsection*{Mesh strategy and $y^+$ criterion}

A predominantly structured mesh with local refinement was applied in regions of high velocity gradients and near-wall flow. The standard $k–\epsilon$ turbulence model with wall functions was used; therefore, the mesh was designed to meet the wall treatment requirements, targeting 30 < $y^+$ < 300 on most wall surfaces to ensure accurate and computationally efficient boundary layer modeling. The thickness of the first layer y = $10^{-4}$ m.

\subsection*{Solver settings}

The coupled Joule--thermal--flow problem is iterated to convergence with a relative residual criterion of $10^{-3}$ on all fields. The simulations are steady-state, so no time stepping is involved. The linear system is solved with the COMSOL Multiphysics linear iterative segregated solver. The damping factors are 1 for electric field, 0.5 for velocity and pressure, 0.5 for temperature, 0.35 for turbulence variable.

\clearpage
\section{FLiBe-coolant properties used in processing experimental data and developing a validation model \emph{FE-ref}.}

\begin{table}[ht]
\centering
\caption{Temperature-dependent thermophysical properties of base FLiBe (66~mol~\% -- 34~mol~\%) for \emph{FE-ref} model.}
\label{tab:properties}

\renewcommand{\arraystretch}{1.3}

\begin{tabular}{|c|c|c|c|c|}
\hline
\textbf{Property} &
\textbf{Units} &
$\mathbf{f(T)}$ &
$\mathbf{T,\,K}$ &
\textbf{Data source} \\
\hline

Thermal conductivity ($k$) &
W/(m$\cdot$K) &
$0.00065\,T+0.45$ &
890--1023 &
\cite{bobrova2023thermophysical} \\
\hline

Density ($\rho$) &
g/m$^{3}$ &
$(2.39-0.00047\,T)$ &
890--1023 &
\cite{redkin2021density} \\
\hline

Viscosity ($\eta$) &
Pa$\cdot$s &
\begin{tabular}[c]{@{}c@{}}$-2.44\times10^{-10}T^{3}+2.44\times10^{-7}T^{2}$\\ $-2.44\times10^{-4}T+0.32$\end{tabular} &
890--1023 &
\cite{tkacheva2024current} \\
\hline

Electrical conductivity ($\sigma$) &
S/m &
$\displaystyle
\frac{1}{\exp\!\left(2.36\times10^{3}/T-8.101\right)}
$ &
750--950 &
\cite{edwards1952electrical} \\
\hline

\end{tabular}
\end{table}

\begin{table}[ht]
\centering
\caption{Temperature-dependent thermal conductivity of AISI 316L pipeline.}
\label{tab:AISI_k}

\renewcommand{\arraystretch}{1.3}

\begin{tabular}{|c|c|}
\hline
\textbf{$T$, K} &
\textbf{$k$, W/(m$\cdot$K)} \\
\hline

500 &
16.61 \\
\hline

600 &
18.32 \\
\hline

700 &
19.98 \\
\hline

800 &
21.63 \\
\hline

900 &
23.21 \\
\hline

1000 &
24.68 \\
\hline

1100 &
26.17 \\
\hline

1200 &
27.53 \\
\hline

1300 &
28.96 \\
\hline

\end{tabular}
\end{table}

\begin{table}[ht]
\centering
\caption{Temperature-dependent electrical conductivity of AISI 316L pipeline.}
\label{tab:AISI_sigma}

\renewcommand{\arraystretch}{1.3}

\begin{tabular}{|c|c|}
\hline
\textbf{$T$, K} &
\textbf{$\sigma\cdot 10^{-6}$, S/m} \\
\hline

293 &
1.41 \\
\hline

373 &
1.21 \\
\hline

473 &
1.052 \\
\hline

673 &
0.85 \\
\hline

873 &
0.73 \\
\hline

1073 &
0.55 \\
\hline

\end{tabular}
\end{table}

\clearpage
\section{Errors of MTP for FLiBe-UF$_4$ and FLiBe-LaF$_3$ systems}

\begin{figure}[h]
	\centering 
	\includegraphics[width=1\linewidth]{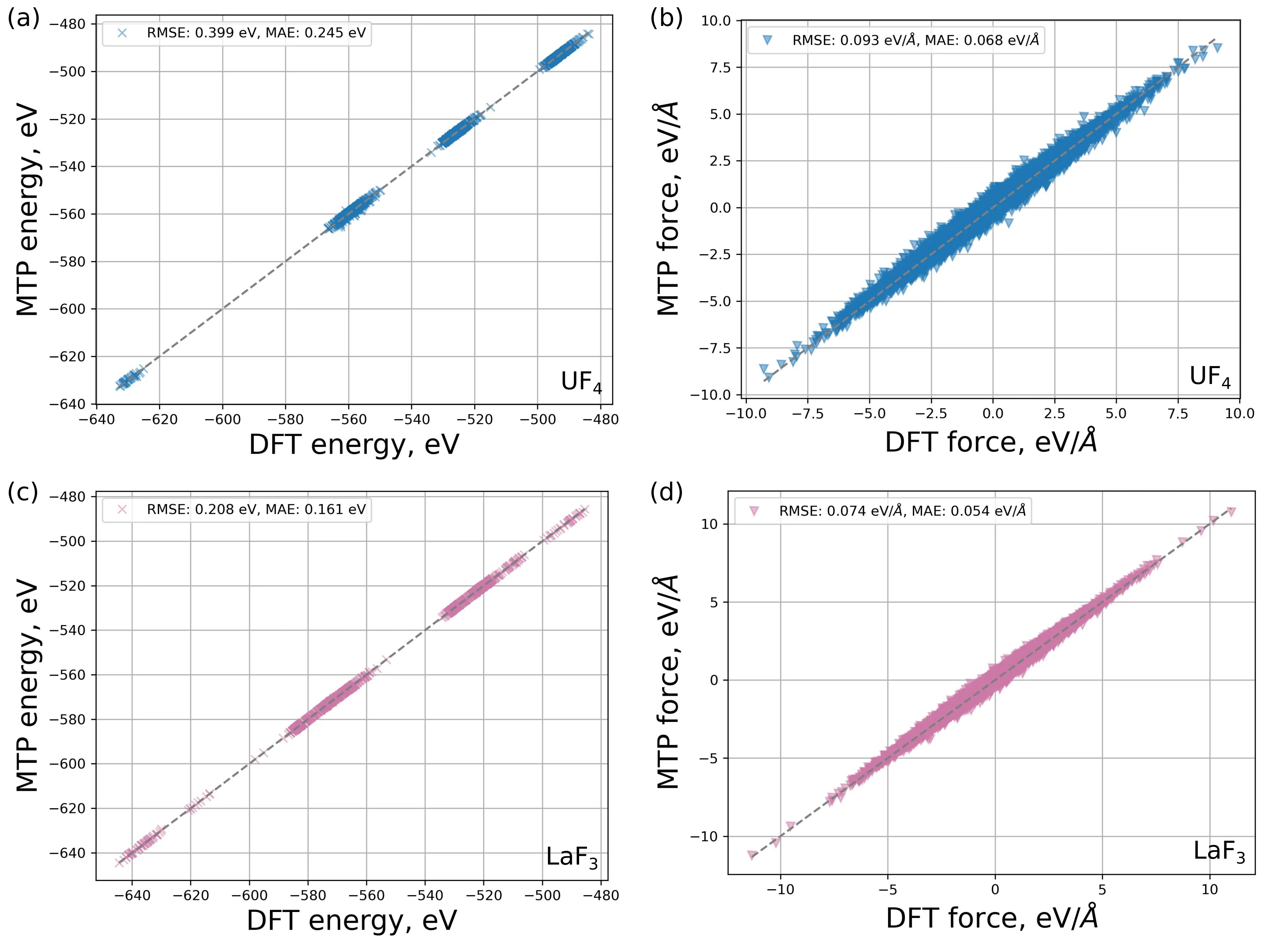}	
	\caption{Errors of MTP on the training set. (a) Energies for FLiBe-UF$_4$, (b) forces acting on atoms for FLiBe-UF$_4$, (c) energies for FLiBe-LaF$_3$, (d) forces acting on atoms for FLiBe-LaF$_3$.} 
	\label{SIfig:errors}
\end{figure}

The correlation plots for energies and forces acting on atoms on the training set are reported in Fig.~\ref{SIfig:errors}. For the FLiBe-UF$_4$ and FLiBe-LaF$_3$ systems we have 1088 and 948 atomic configurations in training set, respectively. The errors in form of root mean sqaure error (RMSE) and mean absolute error (MAE) are reported on plots.

\clearpage
\section{Viscosity calculations autocorrelation functions}

\begin{figure}[h]
	\centering 
	\includegraphics[width=1\linewidth]{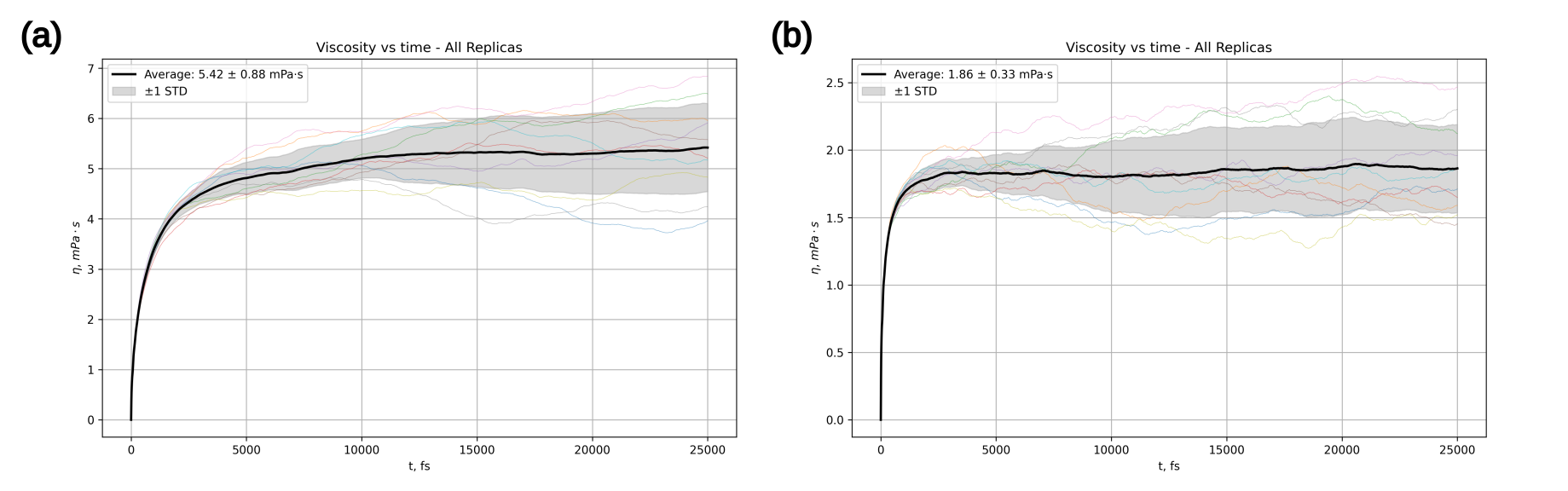}	
	\caption{Stress-stress autocorrelation function integrals for the FLiBe 66~mol~\% -- 34~mol~\% with 5~mol~\% of UF$_4$, obtained at temperature of 973~K (a) and 1273~K (b).} 
	\label{SIfig:ACF}
\end{figure}

In Fig.~\ref{SIfig:ACF} we represent the convergence of viscosity with respect to the stress-stress autocorrelation function. We observe that by 15~ps the value of viscosity is converged. For the higher temperatures the convergence is reached earlier.

\clearpage
\section{Specific heat capacity of FLiBe-UF$_4$ and FLiBe-LaF$_3$}

\begin{figure}[h]
	\centering 
	\includegraphics[width=0.7\linewidth]{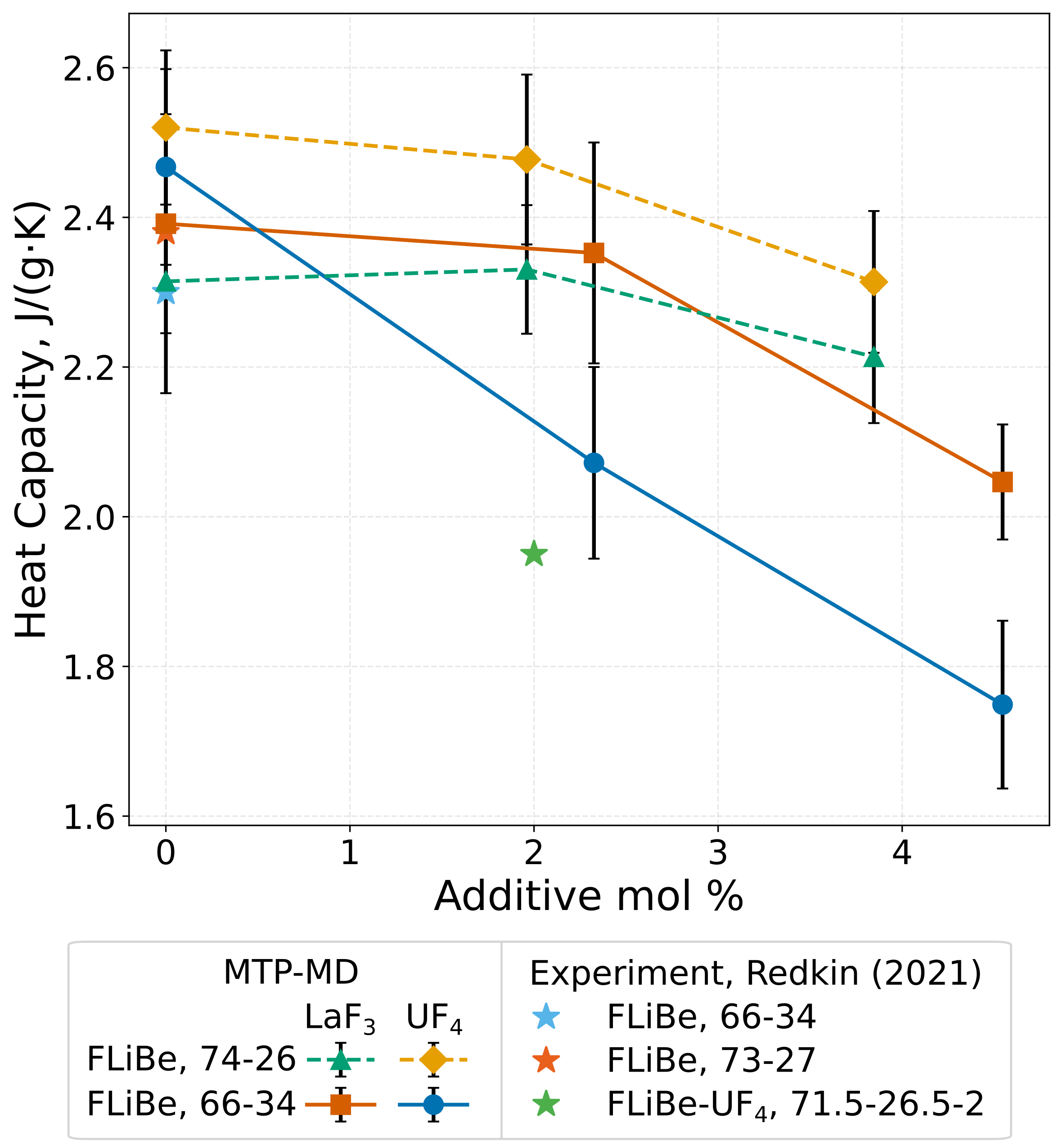}	
	\caption{Composition dependence of the mass-specific heat capacity of molten salts, obtained with MTP-MD. Experimental data for pure FLiBe is taken from Redkin (2021) \cite{redkin2021density}.} 
	\label{SIfig:Cp}
\end{figure}

The specific heat capacity at constant pressure (C$_p$) is obtained from the temperature dependence of the total enthalpy over the range from 873 to 1073 K. In Fig.~\ref{SIfig:Cp} we report the C$_p$ calculated with the MTP-MD. We observe the decrease of C$_p$ with addition of LaF$_3$ and UF$_4$. The C$_p$ for the pure FLiBe 66~mol~\% -- 34~mol~\% is in agreement with the experimental study \cite{redkin2021density}.   

\clearpage

\section{Diffusion coefficients of FLiBe 74~mol~\% -- 26~mol~\% with UF$_4$ and LaF$_3$}

The diffusion coefficients of species in FLiBe (74~mol~\% -- 26~mol~\%) show the similar behavior to one discussed for the FLiBe (66~mol~\% -- 34~mol~\%) in the main text.

\begin{figure}[h]
	\centering 
	\includegraphics[trim={0cm 0cm 0cm 0cm}, clip, width=1\linewidth]{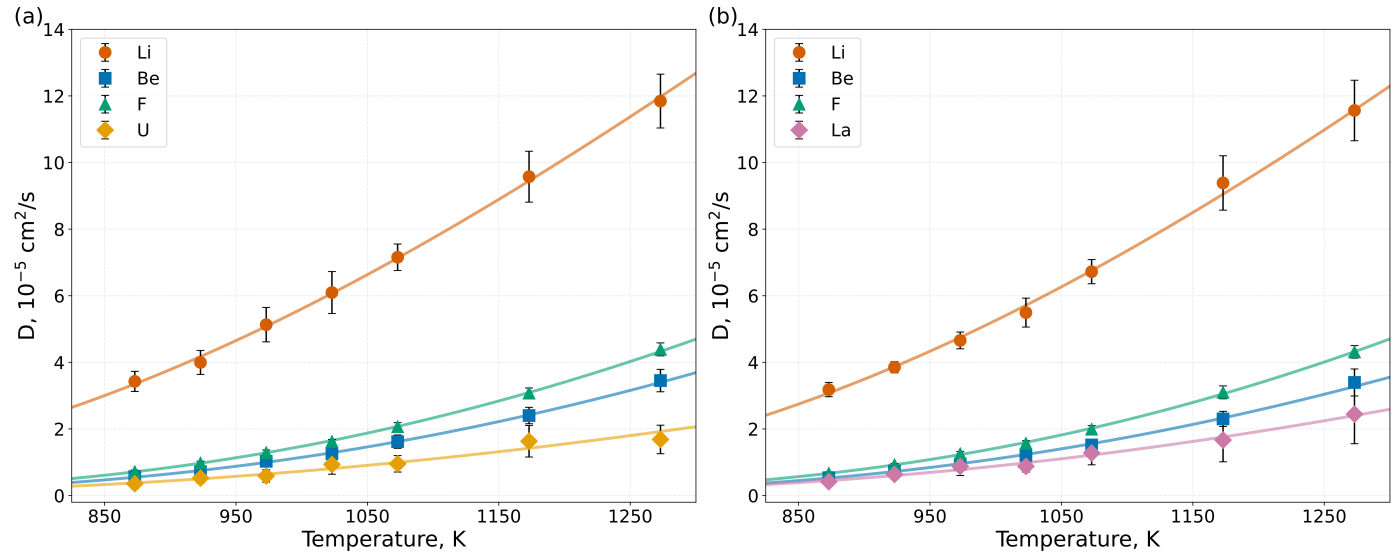}	
	\caption{Diffusion coefficients of species in FLiBe (74~mol~\% -- 26~mol~\%) with 4~mol~\% additions of UF$_4$ (a) and LaF$_3$ (b).} 
	\label{SIfig:diffusivity}
\end{figure}

\end{document}